\documentclass[aps,prd,floatfix,superscriptaddress,amsmath,amssymb,11pt]{revtex4-2}
\usepackage{amsfonts}
\usepackage{graphicx,color}
\usepackage{times}
\usepackage{amsmath} 

\usepackage{booktabs}

\usepackage{multirow} 

\usepackage{tabularx} 

\usepackage{makecell}
\usepackage{float}

\usepackage{enumitem}

\usepackage{xcolor} 
\usepackage{hyperref} 

\begin{document}
\title{Accelerated Bayesian Inference for Pulsar Timing Arrays: Normalizing Flows for Rapid Model Comparison Across Stochastic Gravitational-Wave Background Sources}

\author{Junrong Lai}
\author{Changhong Li}
\email{changhongli@ynu.edu.cn}
\affiliation{Department of Astronomy,  Key Laboratory of Astroparticle Physics of Yunnan Province, School of Physics and Astronomy,  Yunnan University, No.2 Cuihu North Road, Kunming, China 650091}

\begin{abstract}
The recent detection of nanohertz stochastic gravitational-wave backgrounds (SGWBs) by pulsar timing arrays (PTAs) promises unique insights into astrophysical and cosmological origins. However, traditional Markov Chain Monte Carlo (MCMC) approaches become prohibitively expensive for large datasets. We employ a normalizing flow (NF)-based machine learning framework to accelerate Bayesian inference in PTA analyses. For the first time, we perform Bayesian model comparison across SGWB source models in the framework of machine learning  by training NF architectures on the PTA  dataset (NANOGrav 15-year) and enabling direct evidence estimation via learned harmonic mean estimators. Our examples include 10 conventional SGWB source models such as supermassive black hole binaries, power-law spectrum,  cosmic strings, domain walls, scalar-induced GWs, first-order phase transitions, and dual scenario/inflationary gravitational wave. Our approach jointly infers 20 red noise parameters (10 pulsars) and 2 SGWB parameters per model in $\sim 20$\,hours (including training), compared to $\sim 10$\,days with MCMC (68 pulsars). Critically, the NF method preserves rigorous model selection accuracy, with small Hellinger distances ($\lesssim 0.3$) relative to MCMC posteriors, and reproduces MCMC-based Bayes factors across all tested scenarios. This scalable technique for SGWB source comparison will be essential for future PTA expansions and next-generation arrays such as the SKA, may offer substantial efficiency gains without sacrificing physical interpretability.

\end{abstract}

\date{\today}

\pacs{}
\maketitle

\section{Introduction}

Pulsar timing arrays (PTAs)---including NANOGrav~\cite{NANOGrav:2023hde}, EPTA~\cite{EPTA:2023fyk}, PPTA~\cite{Zic:2023gta}, IPTA~\cite{IPTADR2}, and CPTA~\cite{Xu:2023wog}---have reached unprecedented timing precision, enabling detection of a stochastic gravitational-wave background (SGWB) through spatially correlated fluctuations in pulsar timing residuals. A key hallmark of such detection is the Hellings-Downs (HD) correlation~\cite{Hellings:1983fr}, recently reported by multiple PTA collaborations~\cite{NANOGrav:2023gor,EPTA:2023fyk,Reardon:2023gzh,Antoniadis:2022pcn,Xu:2023wog}.

Theoretical models for SGWB generation span a wide landscape, including mergers of supermassive black hole binaries (SMBHBs), first-order phase transitions (FOPT), cosmic strings, domain walls, scalar-induced GWs, and inflationary/bouncing universe scenarios (see~Appendix~\ref{sec:modeldescription} for details of these models). Discriminating among these possibilities requires Bayesian inference on a growing number of high-dimensional parameters across diverse spectral shapes. However, traditional Bayesian tools---such as Markov Chain Monte Carlo (MCMC) and nested sampling algorithms~\cite{NANOGrav:2023hvm,Bian:2023dnv,Gouttenoire:2023bqy}---have become computationally prohibitive for large datasets like the NANOGrav 15-year release (NG15), especially when extensive model comparison is required.

To address this challenge, we build on recent developments in machine learning by implementing a normalizing flow (NF)-based Bayesian inference pipeline~\cite{Shih:2023jme, Vallisneri:2024xfk, Srinivasan:2024uax}. Our architecture is trained on forward-simulated pulsar timing residuals for multiple SGWB+noise models, including realistic HD correlations and red noise components (see Appendix~\ref{sec:workflow} for the workflow of our training). The NF model maps between the parameter space and a uniform latent distribution via invertible autoregressive flows, enabling efficient posterior reconstruction for each model.

Crucially, we show that this framework not only replicates MCMC-level accuracy for parameter inference, but also enables direct estimation of model evidence through a learned harmonic mean estimator (HME)~\cite{Polanska:2024arc,spurio-mancini:harmonic_sbi,harmonic}. Applied to 10-pulsar subsets of NG15 data, our pipeline yields robust posterior distributions and Bayes factors for ten SGWB source models, including variations of dual inflationary/bouncing universe scenarios. Compared to traditional inference workflows MCMC (68 pulsars), our method reduces runtime 
from  $\sim$ 10 days to $\sim$ 20 hours (10 pulsars) (see Sec.~\ref{sec:summary} for more detailed discussion on the scalability trend, and see Appendix~\ref{sec:time_comparison} for comparative timing of NF and MCMC methods), while maintaining physical interpretability and consistency with MCMC benchmarks (Hellinger distances $\lesssim 0.3$).

Our results demonstrate that NF-based model comparison is a powerful and scalable tool for PTA-era gravitational-wave cosmology. This framework is well-suited for upcoming large-scale datasets from SKA and next-generation PTAs, opening new avenues for rapid inference across the full landscape of SGWB source hypotheses.


\section{Extracting the SGWB Power Spectrum from Pulsar Timing Residuals}

Following standard pulsar timing array (PTA) conventions, each pulsar’s timing residuals can be decomposed into white noise, intrinsic red noise, and a stochastic gravitational-wave background (SGWB) contribution:
\begin{equation}
    r_I(t) = r_I^{\mathrm{WN}}(t) + r_I^{\mathrm{RN}}(t) + r_I^{\mathrm{SGWB}}(t),
\end{equation}
where \(I = 1, \dots, N_\mathrm{pulsars}\) labels the \(I\)-th pulsar and \(N_{\mathrm{pulsars}}\) is the total number of pulsars. In this work, from the NANOGrav 15-year (NG15) dataset \cite{ZenodoNG}, we select ten pulsars previously identified as key contributors to SGWB detection sensitivity following \cite{Arzoumanian_2020,Shih:2023jme}, $N_\mathrm{pulsars}=10$. The white noise residuals of these pulsars satisfy \(r_I^{\text{WN}}(t) \sim \mathcal{N}(0, \sigma_I^2)\), with \(\sigma_I^2\) given at the Table~\ref{tab:pulsar_residuals_white_noise} in Appendix~\ref{sec:data}.

A discrete Fourier transform~\cite{Shih:2023jme} approximates these timing residuals as
\begin{equation}
r_I(t) \approx r_I^{\text{WN}}(t) + \sum_{k=0}^{N_f - 1} \Delta f \left[ a_{I}(f_k) \cos(2\pi f_k t) + b_{I}(f_k) \sin(2\pi f_k t) \right],
\label{mock_residuals}
\end{equation}
with $\langle a_I(f) b_J(f') \rangle = 0$,
\begin{equation}
\langle a_I(f) a_J(f') \rangle = \langle b_I(f) b_J(f') \rangle = S_{IJ}(f)\,\delta(f - f'),
\label{fourier}
\end{equation}
and both $a_I$ and $b_I$ are drawn from the Gaussian distributions, 
where the red noise \(r_I^{\mathrm{RN}}(t)\) and SGWB \(r_I^{\mathrm{SGWB}}(t)\) are captured by a single power spectral density (PSD) matrix,
\begin{equation}
S_{IJ}(f) = S_{IJ}^{\text{RN}}(f) + S_{IJ}^{\text{SGWB}}(f).
\label{PSD_matrix}
\end{equation}
Here, \(f_k = f_L + k\,\Delta f\), \(f_L = \Delta f = 1/T_{\text{obs}}\), and \(T_{\text{obs}} \approx 15.8\,\mathrm{yr}\) for NANOGrav's 15-year data, with \(N_f = 14\) frequency bins. The PSD matrix includes pulsar-specific red noise (diagonal entries),
\begin{equation}
    S_{\mathrm{RN},IJ}^{(I)}(f) 
    = \frac{A_{\mathrm{RN}}^{(I)2}}{12\pi^2}
      \left(\frac{f}{f_{\mathrm{yr}}}\right)^{-\gamma_{\mathrm{RN}}^{(I)}}
      f_{\mathrm{yr}}^{-3} \,\delta_{IJ},
\end{equation}
and an SGWB term reflecting inter-pulsar correlations (off-diagonal entries),
\begin{equation}
    S_{\text{SGWB},IJ}(f) 
    = \frac{1}{12\pi^2 f^5} \,\frac{3H_{100}^2}{2\pi^2}\,\Omega_{\text{GW}}(f)\, h^2 \,\Gamma_{IJ},
\end{equation}
where \(A_{\mathrm{RN}}^{(I)}\) and \(\gamma_{\mathrm{RN}}^{(I)}\) are red noise parameters, \(f_{\mathrm{yr}} = 1\,\mathrm{yr}^{-1}\), \(H_{100} = 100\,\mathrm{km\,s}^{-1}\,\mathrm{Mpc}^{-1}\), and \(h \approx 0.7\). The SGWB correlations are captured by the Hellings-Downs matrix \(\Gamma_{IJ}\), which depends on angular separations \(\zeta_{IJ}\) between pulsars~\cite{Hellings:1983fr,NANOGrav:2023gor}:
\begin{equation}
\begin{aligned}
\Gamma_{IJ} 
= \frac{3}{2} \biggl[
   \frac{1 + \cos \zeta_{IJ}}{2} \ln\!\left(\!\frac{1 + \cos \zeta_{IJ}}{2}\right)
 - \,\frac{1 - \cos \zeta_{IJ}}{2} \ln\!\left(\!\frac{1 - \cos \zeta_{IJ}}{2}\right)
\biggr] -\frac{1 - \cos \zeta_{IJ}}{4} + \frac{1}{2},
\end{aligned}
\end{equation}
with \(\zeta_{IJ}\) computed via \texttt{ENTERPRISE}.

By fitting the red noise and SGWB parameters through Bayesian inference—using either Markov Chain Monte Carlo or normalizing-flow methods—one can extract the best-fit spectral shape and amplitude of the SGWB. More specifically, once a posterior distribution over the relevant noise and SGWB parameters is obtained, the reconstructed power spectrum can be visualized by plotting $\Omega_{\mathrm{GW}}(f)$ at each posterior sample or by constructing a posterior predictive distribution. Such a reconstruction provides direct insight into the amplitude and spectral shape of the SGWB, thus illuminating its physical origin.

\section{Trainning Normalizing-Flow architecture}

\subsection{Normalizing Flow Model Construction and Training Process}

Training NF architectures aims to optimize the probability density 
\(p_{\phi}(\tilde{\theta}_{Di}^{(j)} \mid \tilde{\mathbf{x}}_i^{(j)}, \mathcal{H}^{(j)})\) for simulated parameter vectors \(\tilde{\theta} = \{\tilde{\theta}_{Di}^{(j)}\}\) under the physical model \(\mathcal{H} = \{\mathcal{H}^{(j)}\}\) (encompassing both noise and SGWB). Here, \(\tilde{\mathbf{x}} = \{\tilde{\mathbf{x}}_i^{(j)}\}\) denotes the simulated timing residuals, with \(i = 1,\dots,2 \times 10^5\) (the size of the training set), \(j = 1,\dots,10\) (the number of SGWB models in this study), \(D = 22\) (the dimensionality of each SGWB+noise parameter set, comprising 20 noise parameters ($2N_\mathrm{pulsars}$) plus 2 SGWB parameters), \(\phi\) denotes the weight parameters of this NF-based machine learning model (the weight parameter file is saved after each training iteration, and posterior sampling is performed using this file along with the model script after convergence, $\phi\rightarrow\phi_\mathrm{best}$.). The workflow of the NF-based machine learning pipeline for SGWB analysis in this study is illustrated in Fig.~\ref{fig:processFIG} of Appendix~\ref{sec:workflow}, which outlines data extraction from the NG15 dataset, generation of simulated dataset, NF model training,  posterior inference of observational data, Bayes factors computation and SGWB model comparisons.

In NF-based ML, \(p_{\phi}(\tilde{\theta}_{Di}^{(j)} \mid \tilde{\mathbf{x}}_i^{(j)}, \mathcal{H}^{(j)})\) is mapped from a uniform base distribution $p_{\mathrm{base}}(\tilde{\textbf{z}}_{Di}^{(j)})=\text{Uniform}[-1,1]$ (as we assume a uniform prior on the SGWB model parameters~\footnote{The relation between the base distribution \(\tilde{\theta}\) and the prior distribution \(\theta\) is given by \cite{PTAflow}:
\begin{equation}
  \tilde{\theta}
  = 2 \,
  \frac{\theta - \theta_{\mathrm{min}}}{\theta_{\mathrm{max}} - \theta_{\mathrm{min}}}
  - 1,
\end{equation}
where \(\theta_{\mathrm{max}}\) and \(\theta_{\mathrm{min}}\) denote the upper and lower bounds of \(\theta\), respectively. For a uniform prior, one obtains a uniform base. For other prior distributions, such as a standard normal, one obtains a corresponding standard normal base.}
) by the Jacobian determinant $\left|\det\left(\partial \tilde{\mathbf{z}}_{Di}^{(j)}/\partial \tilde{\theta}_{Di}^{(j)}\right)\right|$, 
\begin{equation}
\label{eq:flow_prob_simu}
p_{\phi}(\tilde{\theta}_{Di}^{(j)} \mid \tilde{\mathbf{x}}_i^{(j)}, \mathcal{H}^{(j)})
= p_{\mathrm{base}}\left(\tilde{\mathbf{z}}_{Di}^{(j)}\right)
\;\cdot\;
\left|\det\left(\partial \tilde{\mathbf{z}}_{Di}^{(j)}/\partial \tilde{\theta}_{Di}^{(j)}\right)\right|,
\end{equation}
where \(\mathbf{\tilde{z}} \equiv T_{\phi}(\tilde{\theta}; \tilde{\mathbf{x}}, \mathcal{H})\),
with \(T_{\phi}\) an invertible mapping \(T_\phi: \tilde{\theta} \mapsto \tilde{\mathbf{z}}\) built from autoregressive flows and permutation layers~\cite{nflows,paszke2019pytorch, durkan2019neural,germain2015made, papamakarios2017masked}. This mapping transforms the simulated parameter vector \(\tilde{\theta}\) into \(\tilde{\mathbf{z}}\), which follows the base distribution \(p_{\mathrm{base}}(\tilde{\textbf{z}})\). Here, \(\phi\) denotes the machine-learning model’s weight parameters (saved to file after each training iteration). For each training iteration, the Jacobian determinant \(\left|\det\left(\partial \tilde{\mathbf{z}}_{Di}^{(j)}/\partial \tilde{\theta}_{Di}^{(j)}\right)\right|\) tracks the change in probability density under the mapping. 

Before training of $p_{\phi}(\tilde{\theta}_{Di}^{(j)} \mid \tilde{\mathbf{x}}_i^{(j)}, \mathcal{H}^{(j)})$, we firstly use the Python function \texttt{np.random.uniform} to generate simulated parameters \(\tilde{\theta}=\{\tilde{\theta}_{Di}^{(j)}(\mathcal{H}^{(j)})\}\) for each SGWB+noise model $\mathcal{H}^{(j)}$ (sampled  \( 2 \times 10^5 \) parameter points (22D: 20 red noise parameters for 10  pulsars + 2 SGWB parameters)\footnote{Each training epoch takes approximately \(15\,\mathrm{minutes}\) for 22D. For 34D (20 red-noise parameters for 10 pulsars + 14 SGWB parameters for testing), it takes approximately \(15\,\mathrm{minutes}\) and \(30\,\mathrm{seconds}\) per epoch.} from the prior in Table~\ref{tab:prior_rn}) and Table~\ref{tab:prior_sgwb}) in Appendix~\ref{sec:prior}, and saves the sampling results to a file.

Then we use \texttt{get\_rawdata.py} to call \texttt{micropta\_SGWB$(j)$.py} (this script is implemented based on the definitions in Eqs.~\ref{mock_residuals} and varies for different SGWB models) to generate simulated timing residuals (4944-dimensional for the NG15 dataset, see Table \ref{tab:pulsar_residuals_white_noise}.) \(\tilde{\mathbf{x}}=\{\tilde{\mathbf{x}}_i^{(j)}\!\left(\mathbf{d}_{\text{obs}}, \tilde{\theta}_{Di}^{(j)}, \mathcal{H}^{(j)}\right)\}\) from the parameter set \(\tilde{\theta}=\{\tilde{\theta}_{Di}^{(j)}(\mathcal{H}^{(j)})\}\) with the pulsar observational data \(\mathbf{d}_{\text{obs}}\) (including times of arrival and pulsar positions). The input size of \(T_{\phi}\) equals the dimension of the residuals \(\mathbf{x}\), which is 4944 for the selected NG15 dataset. Then we use \texttt{get\_rawdata.py} to split the resulting dataset, \(\tilde{\mathbf{x}}=\{\tilde{\mathbf{x}}_i^{(j)}\!\left(\mathbf{d}_{\text{obs}}, \tilde{\theta}_{Di}^{(j)}, \mathcal{H}^{(j)}\right)\}\), to be the training set and validation set (9:1)~\cite{TrainingData}. Note that in simulation, different SGWB+noise models $\mathcal{H}^{(j)}$ yield different simulated timing residuals $\tilde{\mathbf{x}}$, as the SGWB contribution depends on the specific model.

To perform training, we use these two prepared simulated datasets, \(\left(\{\tilde{\theta}_{Di}^{(j)}\}, \{\tilde{\mathbf{x}}_i^{(j)}\}\right)\) (stored in two separate files, simultaneously loaded during training), to optimize the parameter probability density $p_{\phi}(\tilde{\theta}_{Di}^{(j)} \mid \tilde{\mathbf{x}}_i^{(j)}, \mathcal{H}^{(j)})$ according to Eq.~\eqref{eq:flow_prob_simu}. In particular, during each training iteration, the model loads pairs of simulated parameters \(\tilde{\theta}\) and residuals \(\tilde{\mathbf{x}}\), then updates the weight parameters \(\phi\) to minimize the loss function (the negative log-likelihood), thereby maximizing the model likelihood. Following Ref.~\cite{papamakarios2021normalizing} (Eq.~(14)), the loss function is defined as:
\begin{align}\label{eq:loss}
\nonumber \mathrm{Loss}(\phi) &\equiv -\frac{1}{N} \sum_{i=1}^N 
   \ln p_{\phi}(\tilde{\theta}_{Di}^{(j)} \mid \tilde{\mathbf{x}}_i^{(j)}, \mathcal{H}^{(j)}) \\
 &= -\frac{1}{N} \sum_{i=1}^N \left[ \ln p_{\mathrm{base}}(\tilde{\mathbf{z}}_{Di}^{(j)})
    + \ln \left|\det\!\left(\partial T_{\phi}(\tilde{\theta}_{Di}^{(j)}; \tilde{\mathbf{x}}_i^{(j)},\mathcal{H}^{(j)})/\partial \tilde{\theta}_{Di}^{(j)}\right)\right|\right],
\end{align}
where \(i = 1,\dots,2 \times 10^5\) is the size of the training set as aforementioned, and we have used $\tilde{\mathbf{z}}^{(j)}_{D_i} = T_{\phi}(\tilde{\theta}^{(j)}_{D_i}; \tilde{\mathbf{x}}^{(j)}_i, \mathcal{H}^{(j)})$ with \(T_{\phi}\) being the aforementioned invertible mapping of this training set. During each training iteration,  $\ln p_{\mathrm{base}}(\tilde{\textbf{z}}_{Di}^{(j)})=\text{log-Uniform}[-1,1]$ is constant term while $\ln \left|\det\!\left(\partial T_{\phi}(\tilde{\theta}_{Di}^{(j)}; \tilde{\mathbf{x}}_i^{(j)},\mathcal{H}^{(j)})/\partial \tilde{\theta}_{Di}^{(j)}\right)\right|$ is trainable term. 

To optimize the weight parameters \(\phi\) (minimizing the loss function $\mathrm{Loss}(\phi)$), we run \texttt{Train\_model.py} (from \cite{PTAflow}) to call the files \texttt{models.py} and \texttt{utils.py} to  compute following equations~\cite{papamakarios2021normalizing} within a loop:
\begin{align}\label{eq:loop}
\nonumber \phi^{(0)} &= \phi_{\mathrm{init}}, \\
\nonumber \phi^{(n+1)} &= \phi^{(n)} - \alpha \,\nabla_{\phi}\,\mathrm{Loss}(\phi), \\
\nabla_{\phi}\,\mathrm{Loss}(\phi) &= -\frac{1}{N}\sum_{i=1}^N 
   \nabla_{\phi} \ln \left|\det\!\left(\partial T_{\phi}(\tilde{\theta}_{Di}^{(j)}; \tilde{\mathbf{x}}_i^{(j)},\mathcal{H}^{(j)})/\partial \tilde{\theta}_{Di}^{(j)}\right)\right|,
\end{align}
where \(\alpha\) is the learning rate ($\alpha=1\times 10^{-3}$ is taken as the initial value and, during training, it automatically decreases into a half when the loss stop decreasing), and \(n\) indexes the training iterations (with \(n=0\) referring to the initial state of the weights, \(\phi_{\mathrm{init}}\), which are randomly initialized according to PyTorch defaults and normal 
distributions; in this study, our result is convergent around $n=50$.). the validation set's simulated residual files and corresponding parameter files are employed to recompute Eq.~(\ref{eq:loss}), yielding the training loss values that serve as a standard for evaluating model performance.

Through execution of multiple training iterations, the model updates its weight parameters ($\phi^{(n)}$) and reduces the loss function until convergence, $\phi^{(n)}\rightarrow \phi_\mathrm{best}$ (yielding the lowest loss). Then we can use the optimal weight parameters \(\phi_{\text{best}}\) to compute the optimal invertible mapping \(T_{\phi_{\text{best}}}\), thereby inferring subsequent posterior \(p_{\phi_\mathrm{best}}(\theta^{(j)}|\mathbf{x}_{\text{obs}}, \mathcal{H}^{(j)})\) under real observational residuals \(\mathbf{x}_{\text{obs}}\) with the base distribution \(p_{\mathrm{base}}\left(T_{\phi_{\text{best}}}(\theta^{(j)}; \mathbf{x}_{\text{obs}},\mathcal{H}^{(j)})\right)\) according to Eq.~(\ref{eq:flow_prob_simu}). In Appendix~\ref{sec:epochs}, we visualize how the convengence happens during training.

\subsection{Posterior Parameter Inference of Observational Data with Trained Normalizing Flows}  
\label{sec:posterior_inference}

With the trained model weights $\phi_{\text{best}}$ acquired, we can generate the post-training SGWB+noise parameters samples for $\mathbf{x}_{\text{obs}}$, $\theta_{D_l}^{(j)}$. In particular, we run \texttt{plot.py}, which calls \texttt{models.py} and \texttt{utils.py} (the same modules used during training), to upload the trained model weights $\phi_{\text{best}}$ (automatically saved during the execution of \texttt{Train\_model.py}), observational data $\mathbf{x}_{\text{obs}}$ (NG15 timing residuals extracted from the raw data using \texttt{ENTERPRISE})  and  \( \{\mathbf{z}_l\} \) (\( N = 10^5 \) independent samples following the base distribution \( p_{\mathrm{base}}(\mathbf{z}) \)) to generate post-training inverse mapping \( T_{\phi_{\text{best}}}^{-1} \) (i.e. the post-training SGWB+noise parameters samples for $\mathbf{x}_{\text{obs}}$, $\theta_{D_l}^{(j)}$), 
\begin{equation}\label{eq:sample_generation}
    \theta^{(j)}_{D_l} = T_{\phi_{\text{best}}}^{-1}(\mathbf{z}^{(j)}_l; \mathbf{x}_{\text{obs}},\mathcal{H}^{(j)}), \quad l=1,\dots,N~.
\end{equation}
The invertible mapping \( T_{\phi_\mathrm{best}}: \theta \mapsto \mathbf{z} \) combines autoregressive flows and permutation layers \cite{papamakarios2021normalizing}, enabling rapid inverse transformations for efficient sampling, \( T_{\phi_{\text{best}}}^{-1} \), in this study. More specifically, our approach employing NF-based ML method jointly infers 20 red noise parameters and 2 SGWB parameters per model in $\sim 20$\,hours (mainly due to the training process, whereas the inverse mapping \(T^{-1}_{\phi_\mathrm{best}}\) takes only \(40\) seconds in this case), compared to $\sim 10$\,days with MCMC. For more details of timing comparison between NF and MCMC method, see Appendix~\ref{sec:time_comparison}.

Consequently, using the invertible mapping \( T_{\phi_{\text{best}}} \) ( \( T_{\phi_{\text{best}}}^{-1} \) ), we determine the
posterior distribution for observational data, $\mathbf{x}_{\text{obs}}$ ~\cite{papamakarios2021normalizing}), 
\begin{equation} \label{true_prob}
     p_{\phi_\mathrm{best}}(\theta^{(j)}_{D_l}|\mathbf{x}_{\text{obs}}, \mathcal{H}^{(j)}) =  p_{\mathrm{base}}\left(T_{\phi_{\text{best}}}(\theta^{(j)}_{D_l}; \mathbf{x}_{\text{obs}},\mathcal{H}^{(j)})\right) 
    \cdot \left|\det \left(\frac{\partial T_{\phi_{\text{best}}}(\theta^{(j)}_{D_l}; \mathbf{x}_{\text{obs}},\mathcal{H}^{(j)})}{\partial \theta^{(j)}_{D_l}}\right)\right|~.
\end{equation}

\section{Visualization of posterior distribution from NF-based ML}
\begin{figure}[h!]
    \centering
    \includegraphics[width=0.5\linewidth]{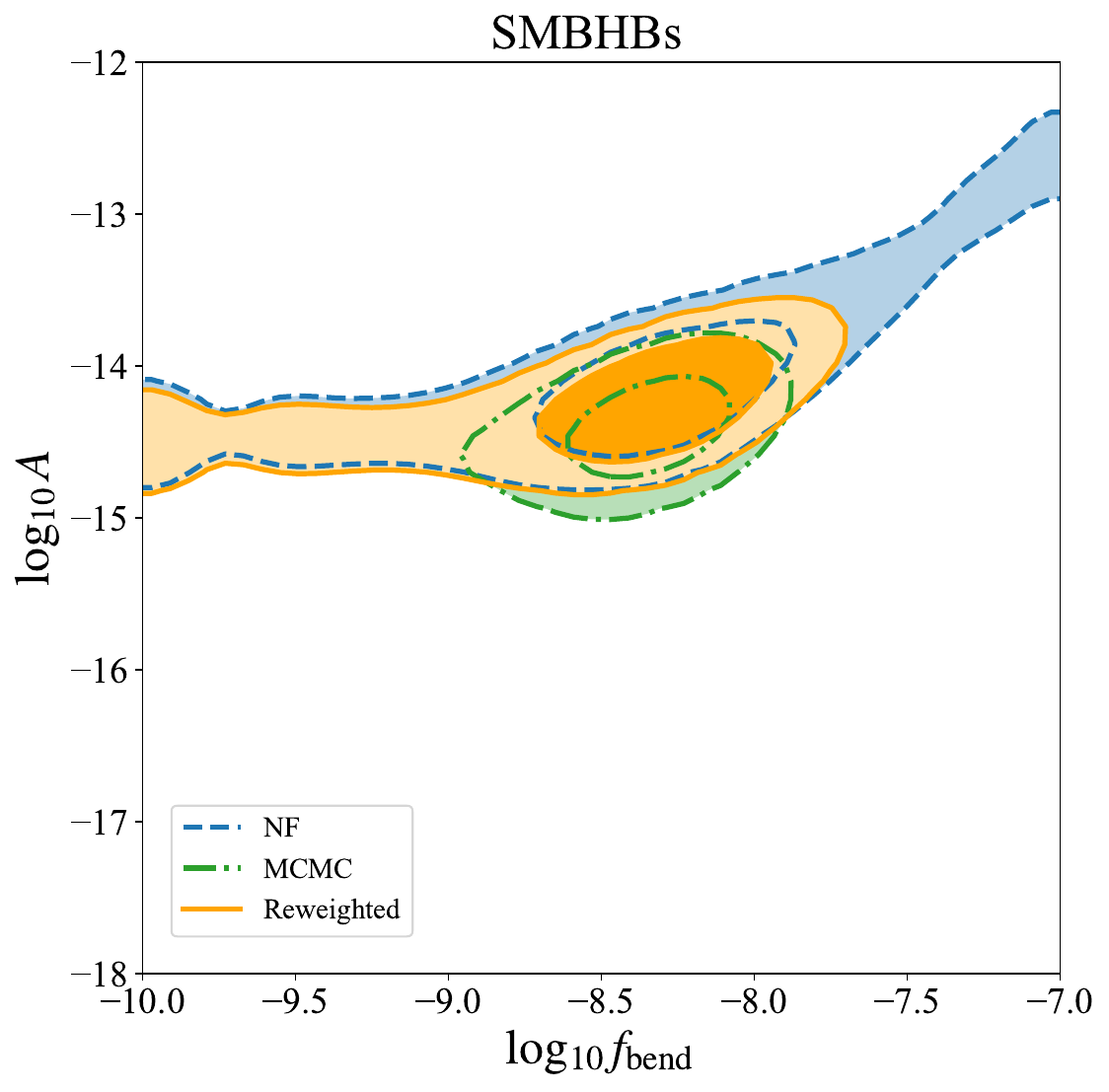}
    \caption{Posterior distributions for SMBHBs in the  \((f_\mathrm{bend}, A)\) plane, where $f_\mathrm{bend}$ is the bending frequency and $A$ is the amplitude. The NF results (the reweighted NF results) are shown as a blue dashed line (orange solid line), while the MCMC results are represented by a green dash-dotted line.}
    \label{fig:A_fbend_RW.pdf}
\end{figure}

\begin{figure}[h!]
    \centering
    \includegraphics[width=0.5\linewidth]{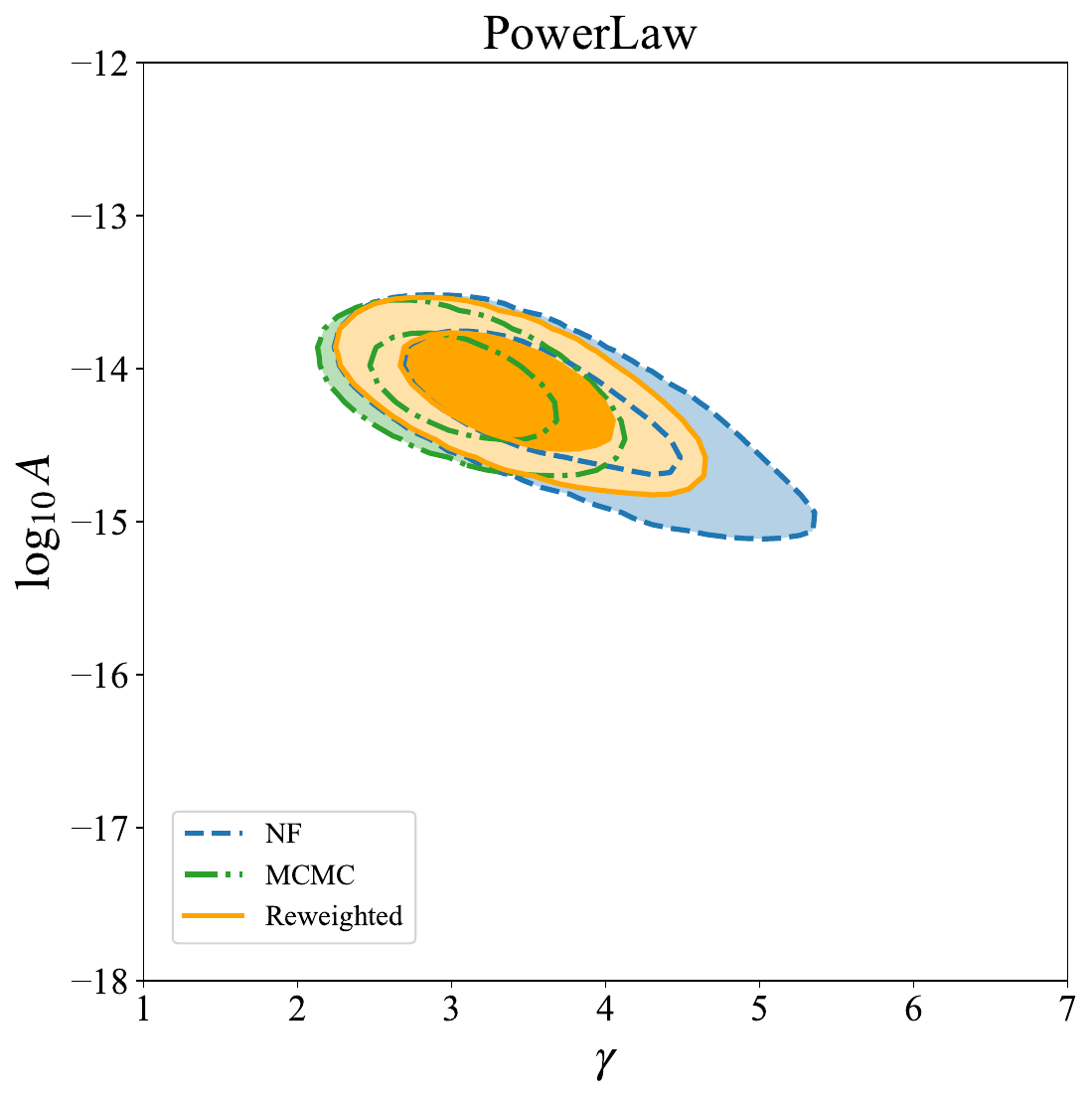}
    \caption{Posterior distributions for Power Law in the \((\gamma, A)\) plane, where $\gamma$ is the spectral index and $A$ is the amplitude. The NF results (the reweighted NF results) are shown as a blue dashed line (orange solid line), while the MCMC results are represented by a green dash-dotted line.}
    \label{fig:A_gamma_RW.pdf}
\end{figure}

\begin{figure}[h!]
    \centering
    \includegraphics[width=0.5\linewidth]{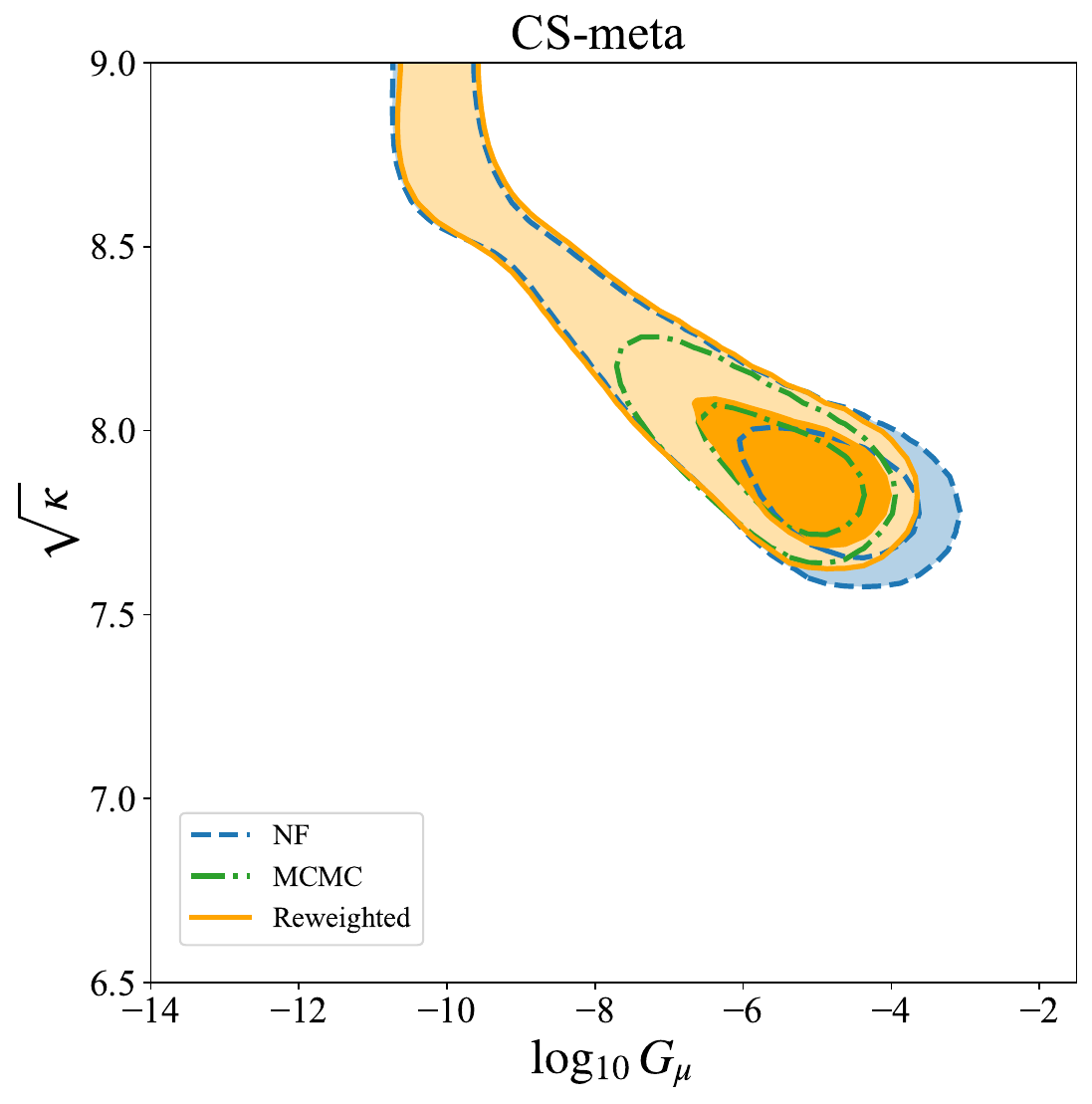}
    \caption{Posterior distributions for Cosmic String-metastable in the \((G\mu, \sqrt{\kappa})\) plane, where $G\mu$ is the string tension and $\sqrt{\kappa}$ is the decay parameters. The NF results (the reweighted NF results) are shown as a blue dashed line (orange solid line), while the MCMC results are represented by a green dash-dotted line.}
    \label{fig:sqrtkappa_Gmu_RW.pdf}
\end{figure}

\begin{figure}[h!]
    \centering
    \includegraphics[width=0.5\linewidth]{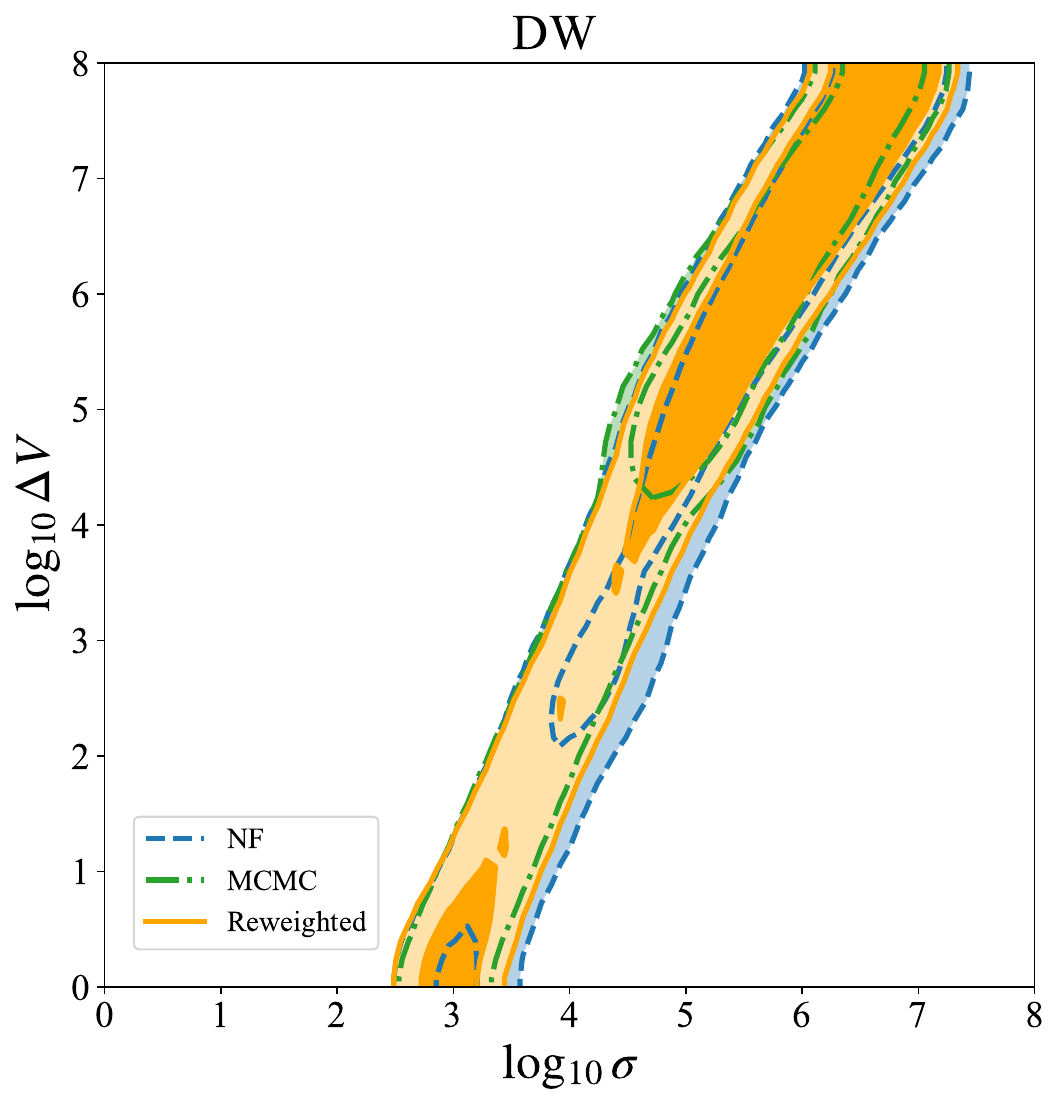}
    \caption{Posterior distributions for Domain Wall in the \((\sigma, \Delta V)\) plane, where $\sigma$ is the domain wall tension and $\Delta V$ is the potential bias. The NF results (the reweighted NF results) are shown as a blue dashed line (orange solid line), while the MCMC results are represented by a green dash-dotted line.}
    \label{fig:DeltaV_sigma_RW.pdf}
\end{figure}

\begin{figure}[h!]
    \centering
    \includegraphics[width=0.5\linewidth]{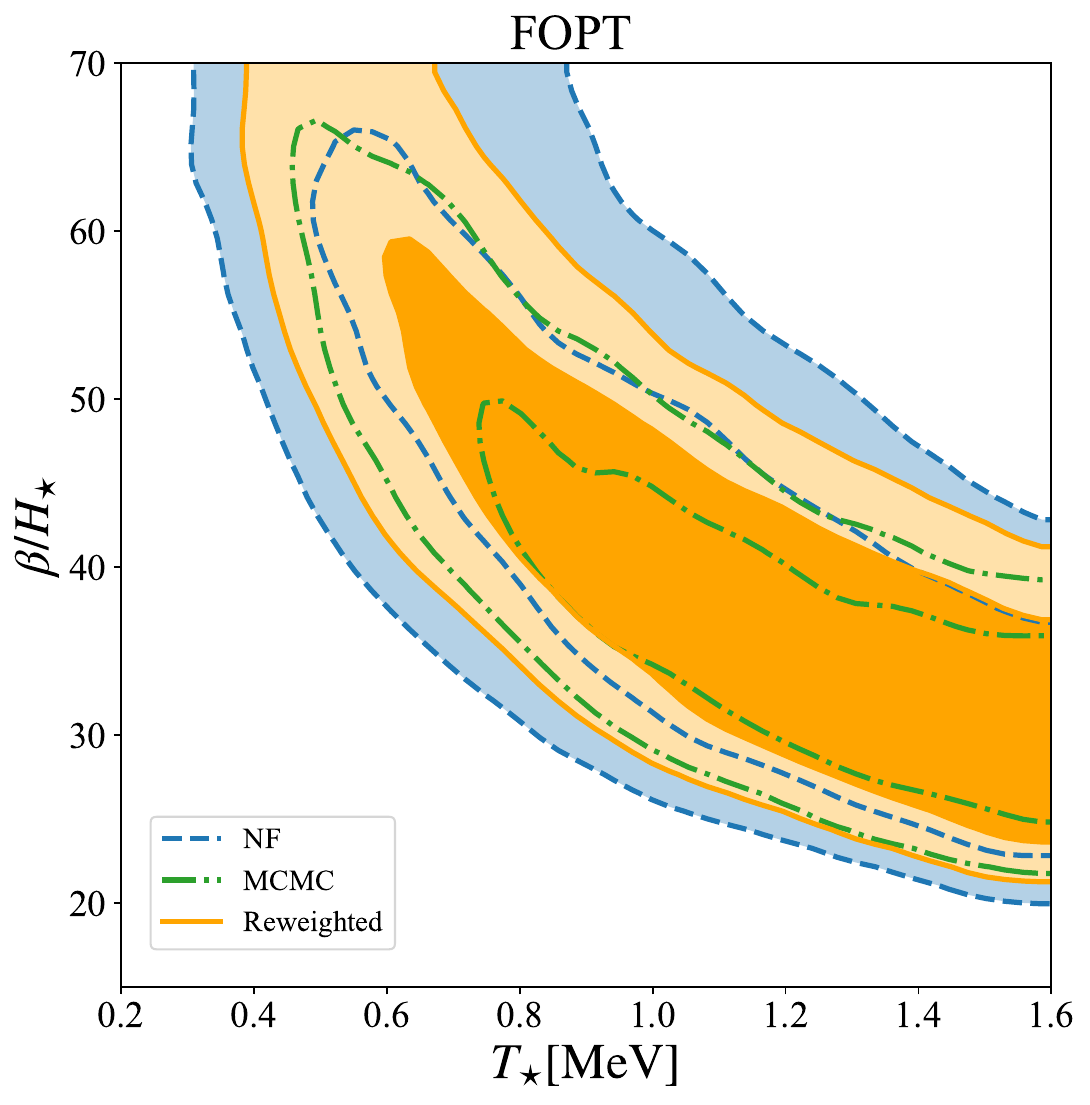}
    \caption{Posterior distributions for FOPT in the \((T_{\star}, \beta/H_{\star})\) plane, where $T_{\star}$ is the temperatures and $\beta/H_{\star}$ is the inverse phase transition durations. The NF results (the reweighted NF results) are shown as a blue dashed line (orange solid line), while the MCMC results are represented by a green dash-dotted line.}
    \label{fig:beta_H_Tstar_RW.pdf}
\end{figure}

\begin{figure}[h!]
    \centering
    \includegraphics[width=0.5\linewidth]{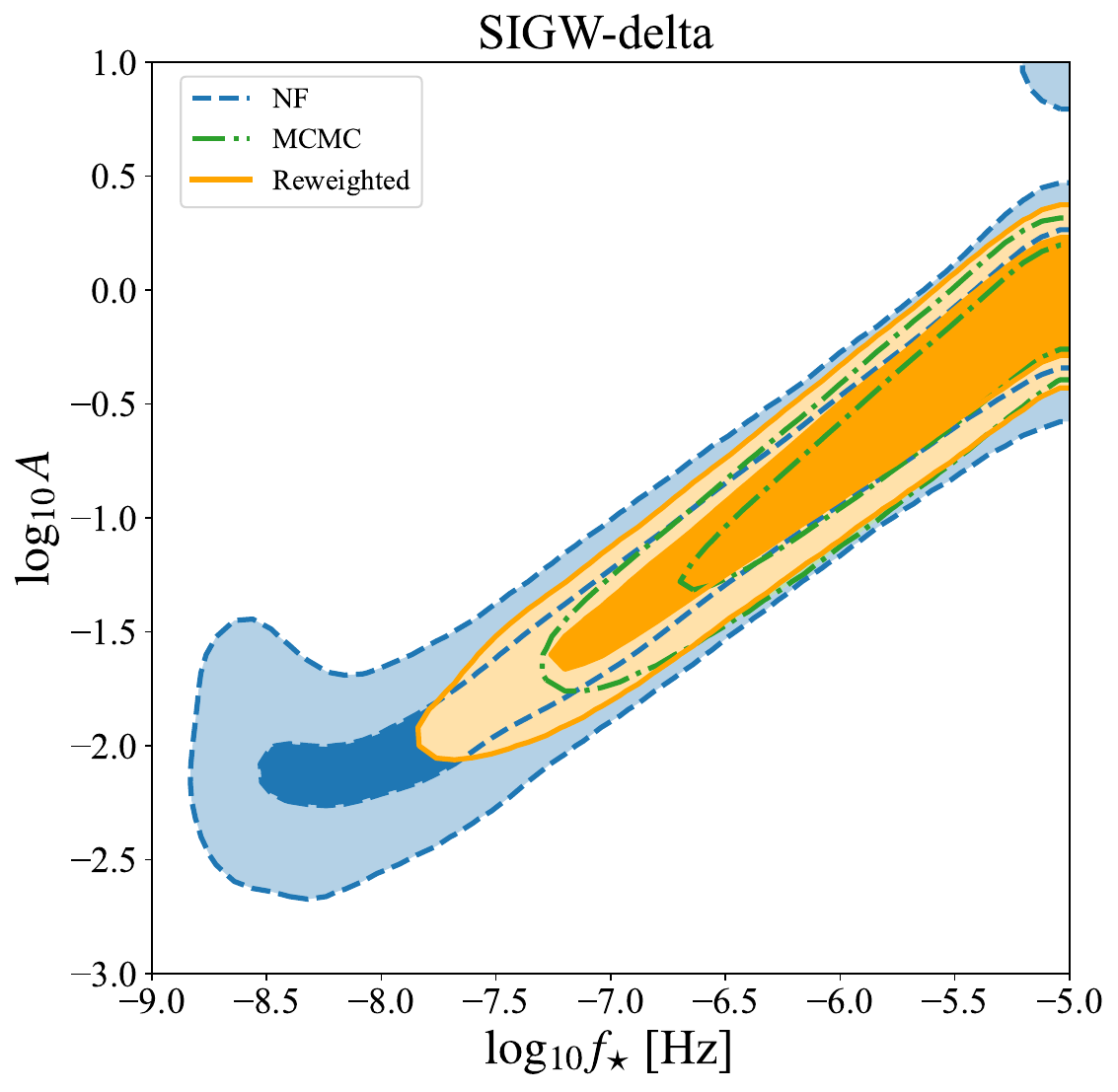}
    \caption{Posterior distributions for SIGW in the \((f_\star, A)\) plane, where $f_\star$ is the temperatures and $A$ is the inverse phase transition durations. The NF results (the reweighted NF results) are shown as a blue dashed line (orange solid line), while the MCMC results are represented by a green dash-dotted line.}
    \label{fig:A_fstar_RW.pdf}
\end{figure}

\begin{figure}[h!]
    \centering
    \includegraphics[width=0.5\linewidth]{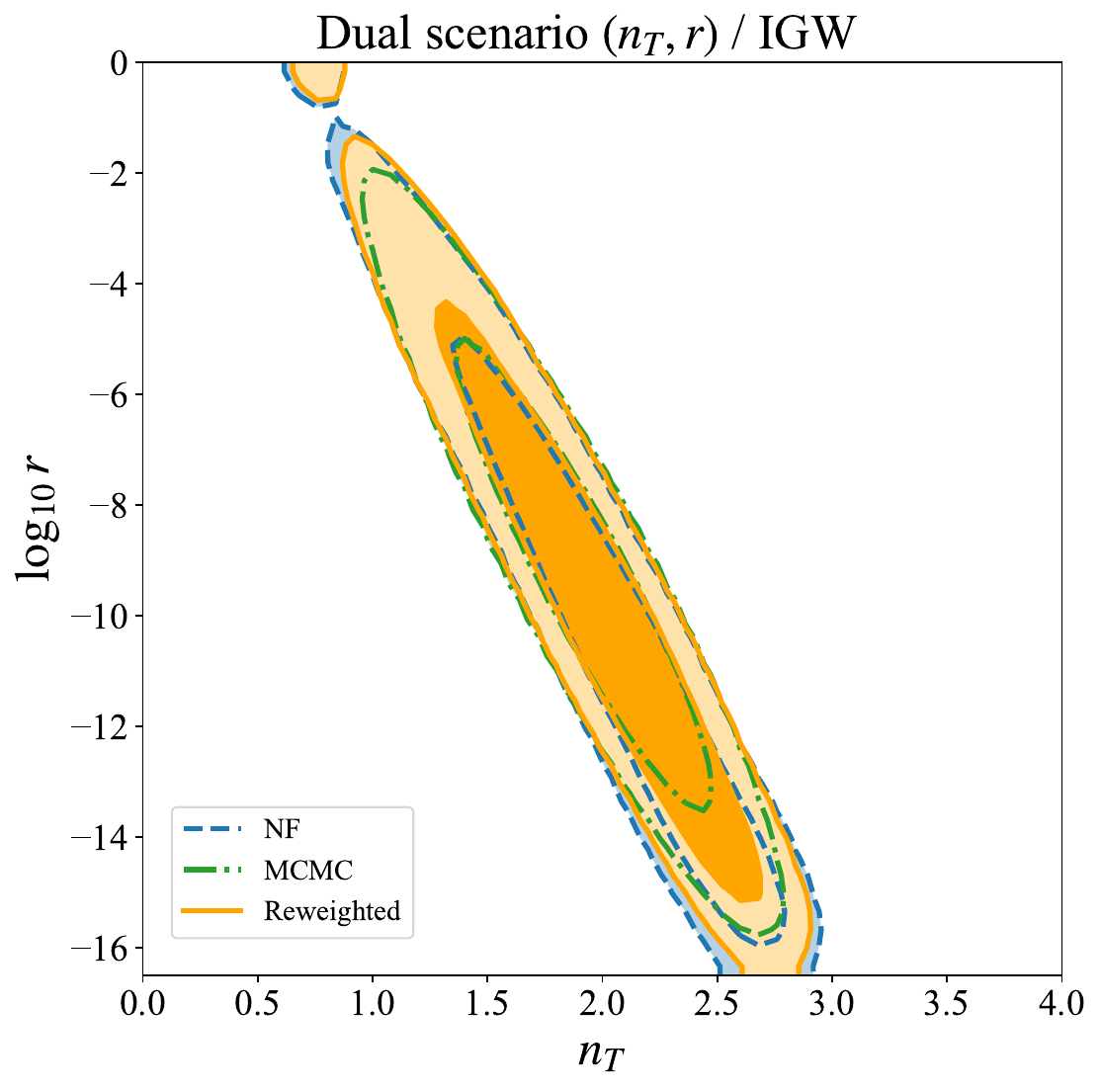}
    \caption{Posterior distributions for the Dual Scenario-$(n_T,r)$/IGW in the \((n_T, r)\) plane, where $n_T$ is the spectral index and $r$ is the tensor-to-scalar ratio. The NF results (the reweighted NF results) are shown as a blue dashed line (orange solid line), while the MCMC results are represented by a green dash-dotted line.}
    \label{fig:r_nT_RW.pdf}
\end{figure}

\begin{figure}[h!]
    \centering
    \includegraphics[width=0.5\linewidth]{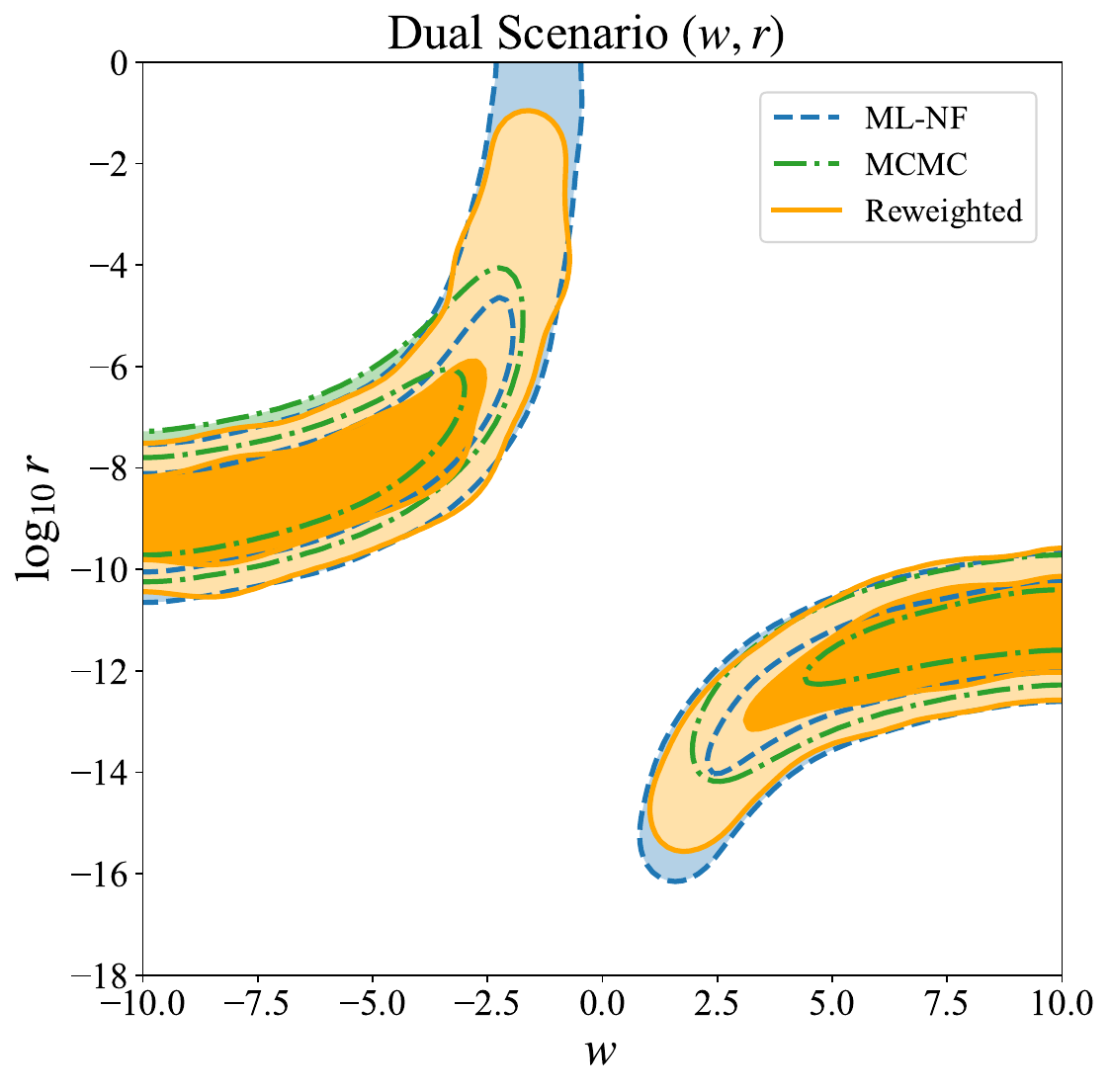}
    \caption{Posterior distributions for the Dual Scenario-$(w,r)$ in the \((w, r)\) plane, where $w$ is the equation of state (EoS) and $r$ is the tensor-to-scalar ratio. The NF results (the reweighted NF results) are shown as a blue dashed line (orange solid line), while the MCMC results are represented by a green dash-dotted line.}
    \label{fig:r_w_RW.pdf}
\end{figure}

\begin{figure}[h!]
    \centering
    \includegraphics[width=0.5\linewidth]{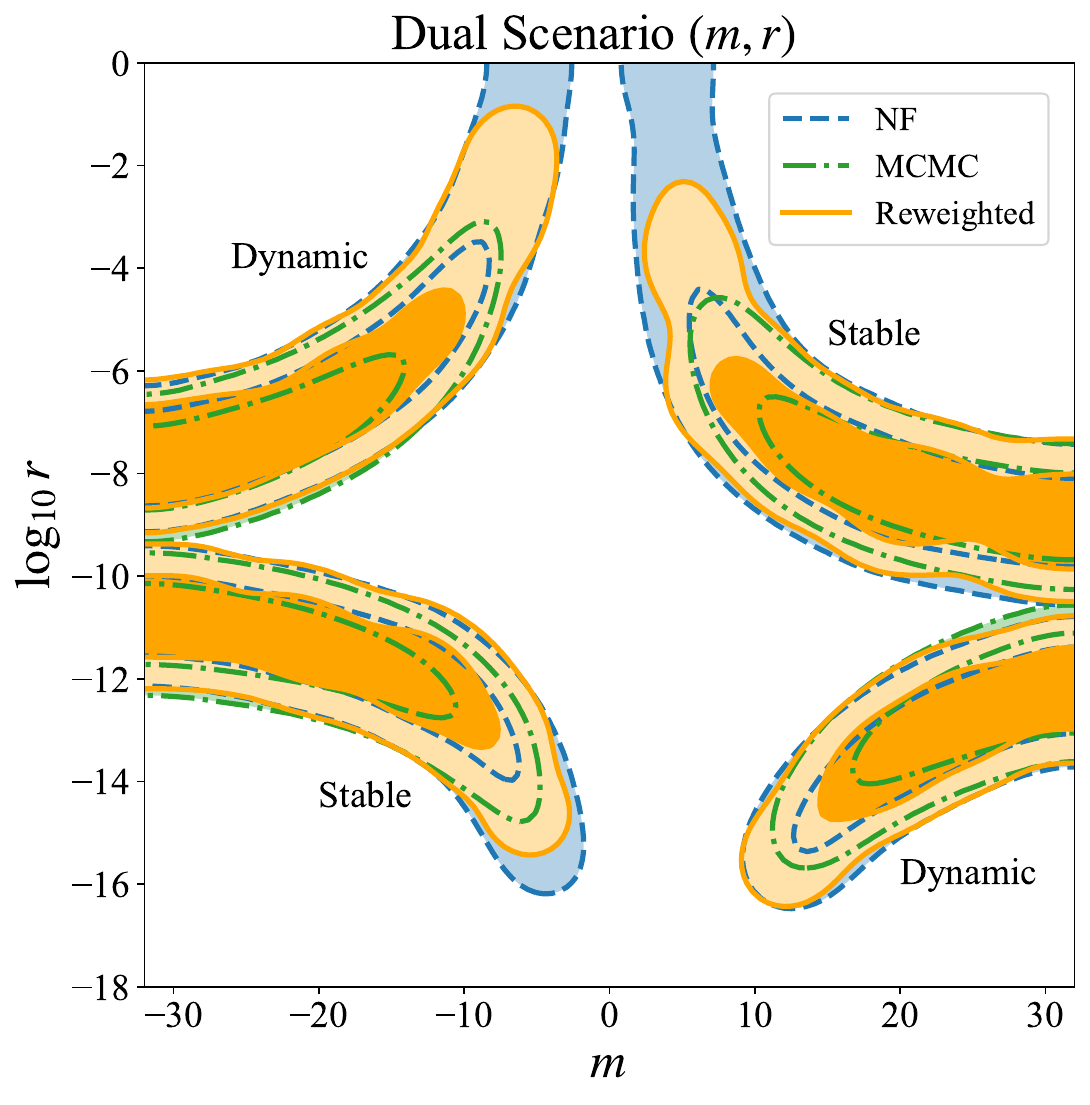}
    \caption{Posterior distributions for the Dual Scenario (Stable+Dynamic solutions) in the \((m, r)\) plane, where \(m\) is the damping parameter and \(r\) is the tensor-to-scalar ratio. The NF results (the reweighted NF results) are shown as a blue dashed line (orange solid line), while the MCMC results are represented by a green dash-dotted line.}
    \label{fig:r_m_RW.pdf}
\end{figure}

To visualize posterior distribution $p_{\phi_\mathrm{best}}(\theta^{(j)}_{D_l}|\mathbf{x}_{\text{obs}}, \mathcal{H}^{(j)})$ for each SGWB source, we decompose the post-training SGWB+noise parameters samples into SGWB part and noise part, 
\begin{equation}
    \{\theta^{(j)}_{D_l}\}= \{\theta^{(j)}_{(D=2)_l, \mathrm{SGWB}}\}+\{\theta^{(j)}_{(D=20)_l, \mathrm{RN}}\}~.
\end{equation} 
At following, we use the script \texttt{plot.py} to plot $p_{\phi_\mathrm{best}}(\theta^{(j)}_{(D=2)_l, \mathrm{SGWB}}|\mathbf{x}_{\text{obs}}, \mathcal{H}^{(j)})$, respectively, for SGWB source models, (1) SMBHBs with environmental effects, (2) PowerLaw,  (3) Cosmic String-metastable (CS-meta),  (4) Domain Walls (DW), (5) FOPT, (6) SIGW-delta, (7) Dual scenario ($(n_T,r)$)/IGW,  (8) Dual scenario ($(w,r)$), (9) (Stable) Dual scenario ($(m,r)$) and (10) (Dynamic) Dual scenario ($(m,r)$), as illustrated in Fig.~\ref{fig:A_fbend_RW.pdf}-Fig.~\ref{fig:r_m_RW.pdf}(Stable+Dynamic). Contours in these figures indicate the 68\% and 95\% credible regions. For the description of each SGWB source model and the prior, see Appendix~\ref{sec:modeldescription} and Appendix~\ref{sec:prior}. For the detailes of the reweighted NF, see Appendix~\ref{sec:reweighted}. And for the detailes of MCMC, see~\cite{Li:2024oru}.

In Fig.~\ref{fig:A_fbend_RW.pdf}-Fig.~\ref{fig:r_m_RW.pdf}, we illustrate the result from the NF and the reweighted NF, and compare them with the result from MCMC (benchmark method). It is evident that they agree well in the high-density regions, indicating that the NF and the reweighted NF methods effectively captures the main parameter constraints from MCMC method for various SGWB source models. In particular, both the 1-\(\sigma\) and 2-\(\sigma\) range from the NF analysis cover (is broader than) the corresponding regions from MCMC, suggesting that the NF method adopts a more conservative coverage in the tails. This behavior is likely related to the chosen training epochs ($n=50$) and number of analyzed pulsars ($N_\mathrm{pulsars}=10$ for NL while $N_\mathrm{pulsars}=68$ for MCMC) in this study (for different epochs ($n=75$), see Fig.~\ref{fig:diff_epochs} of Appendix~\ref{sec:epochs}). To test scalability, we have run epochs for \(N_{\mathrm{pulsars}}=(8,9,10,11,12,13,14,15)\) with number of time residuals data $N_\mathrm{time-residual}=(3797, 4220, 4944, 5604, 6228, 6585, 7063, 7463)$. These results suggest that the per-pulsar processing time - assuming each pulsar has a similar number of time residuals data- may remain approximately constant and that the overall training time appears to grow roughly linearly for $8\le N_{\mathrm{pulsars}}\le 15$. For more detailed discussion on the scalability trend, see Sec.~\ref{sec:summary}. However, these differences do not affect the characterization of the core posterior structure; the central estimates from both methods essentially overlap, demonstrating that the NF approach can achieve comparable accuracy to traditional MCMC while significantly enhancing computational efficiency. 

Furthermore, as described in Appendix~\ref{sec:reweighted}, to achieve more accurate posterior estimates for SGWB sources, these samples directly generated by the NF method are reweighted using the likelihood $\mathcal{L}(\mathbf{x}_{\mathrm{obs}} \mid \theta^{(j)}_{(D=2)_l,\mathrm{SGWB}}, \mathcal{H}^{(j)}) $ \cite{Hourihane:2022ner, Dax:2022pxd, Shih:2023jme}. This reweighting $\{w_l^{(j)}\}$ (see Eq.~(\ref{eq:reweighting}) in Appendix~\ref{sec:reweighted}) increases the sample precision, bringing the resulting distribution closer to the MCMC-derived distribution as illustrated in Fig.~\ref{fig:A_fbend_RW.pdf}-Fig.~\ref{fig:r_m_RW.pdf}. 

Combing these reweighted posterior distributions with the Hellinger distance $H$ calculations in Table~\ref{Hellinger_distances} demonstrates that, after reweighting, the posterior distribution more closely matches the MCMC-derived posterior than the direct NF sampling results. In particular, the Hellinger distance is bounded between 0 and 1, with smaller values indicating closer agreement between the distributions. In practice, \(H < 0.3\) implies that the two distributions are well aligned. Note that \(H < 0.3\) is only an empirical criterion for “well” aligned taken in this study. For more statistical interpretation and astrophysical applications, see Refs.\cite{Devroye1996APT,NANOGrav:2023tcn}. For more details of Hellinger distance calculations see Appendix~\ref{sec:hellinger}.

\begin{table}[h!]
\centering
\footnotesize
\setlength{\tabcolsep}{4pt}
\begin{tabular}{@{}lcc@{}}
\toprule
\textbf{SGWB Model} & \textbf{NF/MCMC} & \textbf{Reweighted/MCMC} \\
\midrule
IGW             & 0.3003 & 0.1239 \\
Dual \((w,r)\)  & 0.3186 & 0.1785 \\
Dual (Stable)   & 0.3555 & 0.1681 \\
Dual (Dynamic)  & 0.2955 & 0.1926 \\
SMBHBs          & 0.5078 & 0.4216 \\
PowerLaw        & 0.4118 & 0.3911 \\
FOPT            & 0.3492 & 0.1797 \\
DW              & 0.2426 & 0.1729 \\
SIGW            & 0.4671 & 0.4554 \\
CSmeta          & 0.4164 & 0.3268 \\
\midrule
\textbf{Mean}   & 0.3665 & 0.2611 \\
\bottomrule
\end{tabular}
\caption{Hellinger distance comparisons for different SGWB spectra: NF versus MCMC, and reweighted NF versus MCMC.}
\label{Hellinger_distances}
\end{table}

\section{Bayes Factor and SGWB source model comparisons}

Model comparison across various SGWB source candidates is pivotal to discriminating and identifying the origin of the nanohertz SGWB signals recently detected by PTAs. In Bayesian inference, one performs model comparison by computing the evidence \(Z^{(j)}\) for each hypothesis \(\mathcal{H}^{(j)}\) and evaluating Bayes factors \(\mathrm{BF}_{ij}\) from the posterior distributions. For two competing models \(\mathcal{H}^{(1)}\) and \(\mathcal{H}^{(2)}\), the Bayes factor is defined as
\begin{equation}
    \mathrm{BF}_{ij} \;=\; Z^{(i)}/Z^{(j)}\,.
\end{equation}
A Bayes factor \(\mathrm{BF}_{12}\gg1\) indicates strong support for  \(\mathcal{H}^{(1)}\) and \(\mathcal{H}^{(2)}\), as listed in Table~\ref{tab:bfint}.
\begin{table}[htbp]
\centering
\label{tab:bfint}
\begin{tabular}{c c}
\toprule
$\mathrm{BF}_{ij}$ & Evidence Strength for $\mathcal{H}^{(i)}$ vs $\mathcal{H}^{(j)}$ \\
\midrule
1--3 & Weak \\
3--20 & Positive \\ 
20--150 & Strong \\
$\geq 150$ & Very strong \\
\bottomrule
\end{tabular}
\caption{Bayes factor interpretation for model comparison. A Bayes factor $B_{ij} = 20$ between candidate model $\mathcal{H}^{(i)}$ and alternative $\mathcal{H}^{(j)}$ corresponds to 95\% confidence in $\mathcal{H}^{(i)}$'s superiority, indicating strong evidence \cite{Bian:2023dnv}.}
\end{table}

In the traditional MCMC framework, Bayes factors are most often obtained via Nested Sampling \cite{Buchner:2021kpm}. However, direct evidence estimation remains challenging in an NF‐based ML pipeline. Here, we overcome this limitation by applying the learned harmonic mean estimator (HME) \cite{Polanska:2024arc,spurio-mancini:harmonic_sbi,harmonic}—an enhanced variant of the classical HME \cite{402e43fa-e885-375b-bdd8-249b07c763b0}—to our NF‐derived posterior samples. This procedure yields the marginal likelihood (evidence) \(Z^{(j)}\) for each SGWB source model, allowing us to compute Bayes factors \(\mathrm{BF}_{ij}\) and, for the first time, perform rigorous model comparison entirely within the NF framework,
\begin{equation}
\frac{1}{Z^{(j)}} = \frac{1}{N} \sum_{i=1}^{N} \frac{\varphi\left(\theta^{(j)}_{(D=2)_l,\text{SGWB}}\right)}{\mathcal{L}\left(\mathbf{x}_\mathrm{obs} | \theta^{(j)}_{(D=2)_l,\,\mathrm{SGWB}},\mathcal{H}^{(j)}\right)\pi\left(\theta^{(j)}_{(D=2)_l,\text{SGWB}}\mid \mathcal{H}^{(j)}\right)},
\end{equation}  
where $\varphi\left(\theta^{(j)}_{(D=2)_l,\text{SGWB}}\right)$ is an arbitrary chosen normalized density introduced to remedy the exploding variance problem of original HME~\cite{10.1111/j.2517-6161.1994.tb01996.x}. Specifically, we employ the Python package \texttt{harmonic} \cite{harmonic} with the two-dimensional SGWB parameter samples \(\theta^{(j)}_{(D=2)_l,\mathrm{SGWB}}\) and their corresponding likelihoods 
\(\mathcal{L}\bigl(\mathbf{x}_{\mathrm{obs}}\mid \theta^{(j)}_{(D=2)_l,\mathrm{SGWB}},\mathcal{H}^{(j)}\bigr)\)
to compute the evidence \(Z^{(j)}\) for each SGWB source model. The Bayes factors \(\mathrm{BF}_{ij} = \frac{Z^{(i)}}{Z^{(j)}}\) are then listed in Table~\ref{tab:bayes_factors_combined_new}. For physical interpretation of Table~\ref{tab:bayes_factors_combined_new}, see Appendix~\ref{sec:phyint}.

For the NF results in Table~\ref{tab:bayes_factors_combined_new}, each model was trained on a dataset of \(2\times10^5\) samples for 50 epochs using identical hyperparameters, and we selected the checkpoint with the lowest loss near epoch 50 for posterior sampling. To reduce the variance of the learned HME, we discarded the lowest 10\% of likelihood values when computing both the log-likelihood and the log-evidence. In the same table, we also report Bayes factors computed from evidence estimates obtained by applying the isocontour integration algorithm of Nested Sampling \cite{Buchner:2021kpm} to a kernel density estimator (KDE) \cite{the_nanograv_collaboration_2023_10344086} of the posterior sampled via MCMC, as implemented in Ceffyl~\cite{Lamb:2023jls}.

\begin{table*}[h]
\label{new_BF_matrix_ML}
\centering
\hspace*{-1.7cm}
\tiny
\setlength{\tabcolsep}{6pt}
\renewcommand{\arraystretch}{1.2}
\begin{tabular}{lcccccccccc}
\toprule
\textbf{MCMC/NF} & \textbf{SMBHB} & \textbf{Powerlaw} & \textbf{CS} & \textbf{DW} & \textbf{FOPT} & \textbf{SIGW} & \textbf{Dual\_nT/IGW} & \textbf{Dual\_w} & \textbf{Dual\_S} & \textbf{Dual\_D} \\ 
\midrule
\textbf{SMBHB}      & 1.0   & 0.6 $\pm$ 0.1   & 1.1 $\pm$ 0.3   & 52.5 $\pm$ 15.5  & 2.7 $\pm$ 0.4   & 0.4 $\pm$ 0.1   & 1.0 $\pm$ 0.2    & 0.2 $\pm$ 0.04   & 0.2 $\pm$ 0.04   & 0.3 $\pm$ 0.1   \\ 
                     & 1.0     & 0.5 $\pm$ 0.01  & 0.8 $\pm$ 0.01  & 55.8 $\pm$ 1.9   & 2.0 $\pm$ 0.03  & 0.4 $\pm$ 0.01 & 0.8 $\pm$ 0.01   & 0.1 $\pm$ 0.002  & 0.2 $\pm$ 0.003  & 0.3 $\pm$ 0.004  \\ \hline
\textbf{Powerlaw}    & 1.7 $\pm$ 0.4   & 1.0   & 1.8 $\pm$ 0.6   & 89.7 $\pm$ 28.9  & 4.6 $\pm$ 0.9   & 0.7 $\pm$ 0.2   & 1.7 $\pm$ 0.5    & 0.3 $\pm$ 0.1   & 0.3 $\pm$ 0.1   & 0.5 $\pm$ 0.1   \\ 
                     & 1.9 $\pm$ 0.03  & 1.0    & 1.6 $\pm$ 0.03  & 106.7 $\pm$ 3.7   & 3.9 $\pm$ 0.1  & 0.7 $\pm$ 0.01 & 1.6 $\pm$ 0.02   & 0.3 $\pm$ 0.004  & 0.3 $\pm$ 0.01  & 0.5 $\pm$ 0.01  \\ \hline
\textbf{CS}      & 0.9 $\pm$ 0.3 & 0.5 $\pm$ 0.2   & 1.0   & 49.0 $\pm$ 18.4  & 2.5 $\pm$ 0.7   & 0.4 $\pm$ 0.1   & 0.9 $\pm$ 0.3    & 0.2 $\pm$ 0.1   & 0.2 $\pm$ 0.1   & 0.3 $\pm$ 0.1   \\ 
                     & 1.2 $\pm$ 0.02  & 0.6 $\pm$ 0.01  & 1.0    & 66.0 $\pm$ 2.3   & 2.4 $\pm$ 0.04  & 0.4 $\pm$ 0.007 & 1.0 $\pm$ 0.01   & 0.2 $\pm$ 0.002  & 0.2 $\pm$ 0.003  & 0.3 $\pm$ 0.004  \\ \hline
\textbf{DW}          & 0.02 $\pm$ 0.01 & 0.01 $\pm$ 0.004   & 0.02 $\pm$ 0.01   & 1.0    & 0.1 $\pm$ 0.01   & 0.01 $\pm$ 0.003   & 0.02 $\pm$ 0.01    & 0.003 $\pm$ 0.001   & 0.004 $\pm$ 0.001   & 0.006 $\pm$ 0.002   \\ 
                     & 0.02 $\pm$ 0.001 & 0.01 $\pm$ 0.0003  & 0.02 $\pm$ 0.001  & 1.0    & 0.04 $\pm$ 0.001 & 0.006 $\pm$ 0.0002 & 0.01 $\pm$ 0.001   & 0.002 $\pm$ 0.0001  & 0.003 $\pm$ 0.0001  & 0.005 $\pm$ 0.0002  \\ \hline
\textbf{FOPT}        & 0.4 $\pm$ 0.1 & 0.2 $\pm$ 0.04   & 0.4 $\pm$ 0.1   & 19.6 $\pm$ 5.5   & 1.0   & 0.2 $\pm$ 0.04   & 0.4 $\pm$ 0.1    & 0.1 $\pm$ 0.01   & 0.1 $\pm$ 0.01   & 0.1 $\pm$ 0.02   \\ 
                     & 0.5 $\pm$ 0.01  & 0.3 $\pm$ 0.004  & 0.4 $\pm$ 0.01  & 27.7 $\pm$ 1.0   & 1.0    & 0.2 $\pm$ 0.003 & 0.4 $\pm$ 0.01   & 0.1 $\pm$ 0.001  & 0.1 $\pm$ 0.001  & 0.1 $\pm$ 0.002  \\ \hline
\textbf{SIGW}        & 2.4 $\pm$ 0.6 & 1.4 $\pm$ 0.4   & 2.6 $\pm$ 0.9   & 125.5 $\pm$ 42.2 & 6.4 $\pm$ 1.4   & 1.0  & 2.5 $\pm$ 0.7    & 0.4 $\pm$ 0.1   & 0.5 $\pm$ 0.1   & 0.7 $\pm$ 0.2   \\ 
                     & 2.8 $\pm$ 0.04  & 1.5 $\pm$ 0.02  & 2.4 $\pm$ 0.04  & 155.2 $\pm$ 5.5  & 5.6 $\pm$ 0.1  & 1.0  & 2.3 $\pm$ 0.04   & 0.4 $\pm$ 0.01  & 0.4 $\pm$ 0.01  & 0.7 $\pm$ 0.01  \\ \hline
\textbf{Dual\_nT/IGW}         & 1.0 $\pm$ 0.2 & 0.6 $\pm$ 0.2   & 1.1 $\pm$ 0.4   & 53.5 $\pm$ 18.0  & 2.7 $\pm$ 0.6   & 0.4 $\pm$ 0.1   & 1.0    & 0.2 $\pm$ 0.04   & 0.2 $\pm$ 0.1   & 0.3 $\pm$ 0.1   \\ 
                     & 1.2 $\pm$ 0.02  & 0.6 $\pm$ 0.01  & 1.0 $\pm$ 0.02  & 67.9 $\pm$ 2.3   & 2.5 $\pm$ 0.03  & 0.4 $\pm$ 0.007 & 1.0    & 0.2 $\pm$ 0.002  & 0.2 $\pm$ 0.003  & 0.3 $\pm$ 0.004  \\ \hline
\textbf{Dual\_w}    & 6.2 $\pm$ 1.4 & 3.6 $\pm$ 0.9   & 6.6 $\pm$ 2.1   & 324.4 $\pm$ 104.5 & 16.6 $\pm$ 3.3  & 2.6 $\pm$ 0.7   & 6.1 $\pm$ 1.7    & 1.0   & 1.2 $\pm$ 0.3   & 1.9 $\pm$ 0.4   \\ 
                     & 7.3 $\pm$ 0.1  & 3.8 $\pm$ 0.06  & 6.1 $\pm$ 0.09  & 405.0 $\pm$ 14.0 & 14.6 $\pm$ 0.2   & 2.6 $\pm$ 0.04 & 6.0 $\pm$ 0.08   & 1.0    & 1.1 $\pm$ 0.02  & 1.8 $\pm$ 0.03  \\ \hline
\textbf{Dual\_S} & 5.3 $\pm$ 1.2 & 3.1 $\pm$ 0.8   & 5.7 $\pm$ 1.8   & 278.1 $\pm$ 89.9  & 14.2 $\pm$ 2.9  & 2.2 $\pm$ 0.6   & 5.2 $\pm$ 1.4    & 0.9 $\pm$ 0.2   & 1.0  & 1.6 $\pm$ 0.4   \\ 
                     & 6.5 $\pm$ 0.1  & 3.4 $\pm$ 0.07  & 5.5 $\pm$ 0.1  & 362.1 $\pm$ 13.1 & 13.1 $\pm$ 0.2 & 2.3 $\pm$ 0.04 & 5.3 $\pm$ 0.1   & 0.9 $\pm$ 0.02  & 1.0    & 1.6 $\pm$ 0.03  \\ \hline
\textbf{Dual\_D} & 3.3 $\pm$ 0.6 & 1.9 $\pm$ 0.4   & 3.5 $\pm$ 1.1   & 172.3 $\pm$ 52.6  & 8.8 $\pm$ 1.5   & 1.4 $\pm$ 0.3   & 3.2 $\pm$ 0.8   & 0.5 $\pm$ 0.1   & 0.6 $\pm$ 0.1   & 1.0  \\ 
                     & 4.0 $\pm$ 0.06  & 2.1 $\pm$ 0.03  & 3.4 $\pm$ 0.1  & 221.8 $\pm$ 7.6  & 8.0 $\pm$ 0.1  & 1.4 $\pm$ 0.02 & 3.3 $\pm$ 0.1  & 0.5 $\pm$ 0.01  & 0.6 $\pm$ 0.01  & 1.0    \\
\bottomrule
\end{tabular}
\caption{Bayes factors (BF) for different models using NG15 data, evaluated via MCMC and posterior of NF. Each entry is the ratio of the row model's evidence to that of the column model. The first row under each model represents the BF via NS, and the second row represents the BF via NF.}
\label{tab:bayes_factors_combined_new}
\end{table*}

Table~\ref{tab:bayes_factors_combined_new} presents a comprehensive comparison of Bayes factors across all SGWB source models (see Appendix~\ref{sec:modeldescription} for model descriptions). In most cases, the NF‐derived Bayes factors agree with those from MCMC, with NF values lying within the uncertainties of traditional nested‐sampler estimates. Only a few models show minor discrepancies, likely due to variations in flow‐model training quality and finite training data. This concordance—together with the Hellinger distances reported in Table~\ref{Hellinger_distances}—demonstrates that rapid SGWB source model comparison can be achieved in an NF‐based ML framework without sacrificing accuracy. Our results pave the way for efficient SGWB source discrimination in future PTA expansions and next‐generation arrays such as the SKA, may offer substantial gains in computational efficiency while preserving physical interpretability.

\section{Summary}
\label{sec:summary}
\begin{figure}[htbp]
    \centering
    \includegraphics[width=0.95\linewidth]{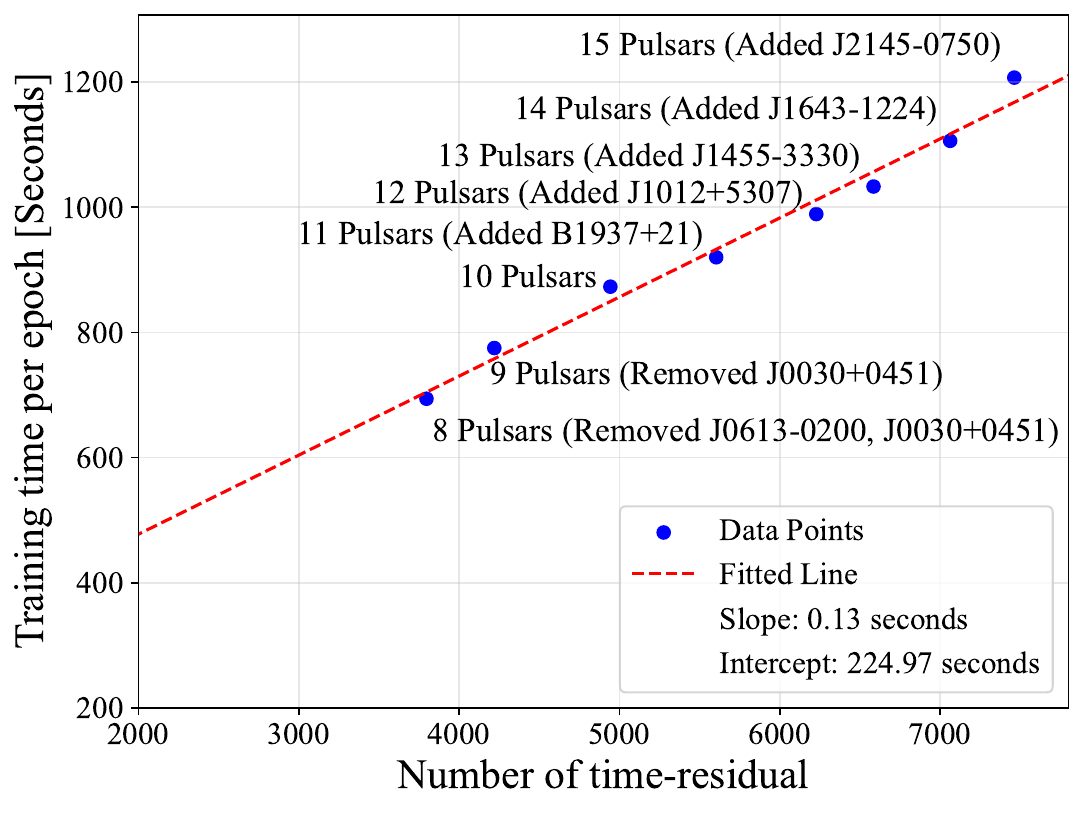}
    \caption{Relationship between training time per epoch and the number of time residuals. The slope of $0.13$ seconds reflects the average time per residual per epoch, \(T_{\mathrm{per-res}}\), while the intercept of $224.97$ seconds captures time for file loading, model initialization, and saving.}
    \label{fig:training_time_vs_residuals}
\end{figure}

In this work, we present a normalizing-flow-based machine learning (NF-based ML) framework for stochastic gravitational-wave background (SGWB) model selection using pulsar timing array data – the first application of ML to SGWB model comparison. In our approach, conditional normalizing flow networks were trained on the NANOGrav 15-year dataset and incorporated a learned harmonic mean estimator to directly infer Bayesian posteriors and model evidences (Bayes factors). We tested ten representative SGWB source models spanning both astrophysical and cosmological scenarios. Despite the high dimensionality (22 parameters per model), the normalized flow-based inference completes in only $\sim$20 hours per model (10 pulsars), compared to roughly $\sim$ 10 days for MCMC analyses (68 pulsars).  Note that the substantial time reduction is partly due to the decrease in dataset size from 68 pulsars to 10 pulsars. It may be also partly attributable to the use of different computers—NF‑ML was run on a GPU computer, while MCMC used a CPU computer—even though the GPU computer for NF‑ML is roughly three times less expensive (See Sec.~\ref{sec:time_comparison} for hardware specifications). In the future, more pulsars will be taken into account. To test scalability, we have run epochs for \(N_{\mathrm{pulsars}}=(8,9,10,11,12,13,14,15)\) with corresponding numbers of time residuals $N_\mathrm{time-residual}=(3797, 4220, 4944,$ $ 5604, 6228, 6585, 7063, 7463)$ with IGW model (Eq.~\eqref{eq:IGWspectrum}). Fitting these points (Fig.~\ref{fig:training_time_vs_residuals}, based on Table~\ref{tab:pulsar_datascale}) yields an average time per residual per epoch of
\begin{equation}
    T_\mathrm{per-res} \simeq 0.13\,\mathrm{s},
\end{equation}
which corresponds to the slope of the fitted line. These results suggest that per‑pulsar processing time—assuming a similar number of residuals per pulsar—remains roughly constant, and that total training time grows nearly linearly. However, we acknowledge that measurements at small \(N_{\mathrm{pulsars}}\) are insufficient to confirm this trend at larger scales. Due to computational resource limits, a full test at \(N_{\mathrm{pulsars}}=68\) (with \(N_\mathrm{time-residual}=20290\)) is not yet feasible, even though the average number of time residuals per pulsar decrease from about 500 to 300. We therefore describe our scalability conjecture as preliminary and plan to extend these studies when additional resources become available.\footnote{We appreciate the anonymous Referee for highlighting this point to us.}

\begin{table}[htbp]
\centering
\begin{tabular}{ccc}
\hline
$N_{\mathrm{pulsars}}$ & $N_{\mathrm{time-residual}}$ & Training time per epoch (Seconds) \\
\hline
8 & 3797 & 694 \\
9 & 4220 & 775 \\
10 & 4944 & 873 \\
11 & 5604 & 920 \\
12 & 6228 & 989 \\
13 & 6585 & 1033 \\
14 & 7063 & 1106 \\
15 & 7463 & 1207 \\
\bottomrule
\end{tabular}
\caption{Summary of Pulsar Data Used in the Scalability Analysis.}
\label{tab:pulsar_datascale}
\end{table}

The posterior distributions obtained with the normalizing flows are in good agreement with those from traditional MCMC sampling, with Hellinger distances typically $\lesssim 0.3$ (on a 0–1 scale where 0 indicates identical distributions). Likewise, the Bayes factors derived from the NF-based ML framework agree with MCMC-based calculations within their reported uncertainties, correctly ranking the evidence for each SGWB model. These findings demonstrate that our ML-driven approach achieves comparable accuracy to standard Bayesian inference while it may significantly reduce runtime by using a GPU setup. In summary, this work provides a robust and faster framework for SGWB model selection, one that is immediately applicable to current PTA datasets and well suited for the demanding analyses of near-future PTA data.

\begin{acknowledgments}
C.L. is supported by the NSFC under Grants No.11963005 and No. 11603018, by Yunnan Provincial Foundation under Grants No.202401AT070459, No.2019FY003005, and No.2016FD006, by Young and Middle-aged Academic and Technical Leaders in Yunnan Province Program, by Yunnan Provincial High level Talent Training Support Plan Youth Top Program, by Yunnan University Donglu Talent Young Scholar, and by the NSFC under Grant No.11847301 and by the Fundamental Research Funds for the Central Universities under Grant No. 2019CDJDWL0005.
\end{acknowledgments}

\begin{appendix}

\section{Data}
\label{sec:data}
We use the NANOGrav 15-year (NG15) wideband dataset \cite{ZenodoNG} and select ten pulsars previously identified as key contributors to SGWB detection sensitivity following \cite{Arzoumanian_2020,Shih:2023jme}. The raw \texttt{.par} and \texttt{.tim} files were processed with \texttt{ENTERPRISE} \cite{enterprise} to extract times of arrival (ToAs), celestial coordinates, white noise parameters (average ToA uncertainties), and timing residuals of these pulsars. Table \ref{tab:pulsar_residuals_white_noise} summarizes each pulsar’s number of timing residuals and corresponding white noise levels.

\begin{table}[h!]
  \centering
  \setlength{\tabcolsep}{10pt}  
  \renewcommand{\arraystretch}{1}  
  \begin{tabular}{l c c}
    \toprule
    Name       & Timing Residuals & White Noise [ns] \\
    \midrule
    J0030+0451 & 724           & 685.7             \\
    J0613-0200 & 423           & 276.0             \\
    J1600-3053 & 481           & 241.7             \\
    J1744-1134 & 433           & 236.3             \\
    J1909-3744 & 833           &  95.4             \\
    J1910+1256 & 216           & 442.1             \\
    J1918-0642 & 487           & 543.2             \\
    J1944+0907 & 180           & 664.4             \\
    J2043+1711 & 459           & 251.4             \\
    J2317+1439 & 708           & 303.6             \\
    Total      & 4944          & –                 \\ 
    \bottomrule
  \end{tabular}
  \caption{Summary of pulsar timing data: Number of timing residuals and average white noise levels (ToA measurement uncertainties) for the ten NG15 pulsars analyzed.}
  \label{tab:pulsar_residuals_white_noise}
\end{table}

\section{Normalizing Flow-Based Machine Learning Training Workflow}
\label{sec:workflow}

\begin{figure}[h!]
    \centering
    \includegraphics[width=0.8\linewidth]{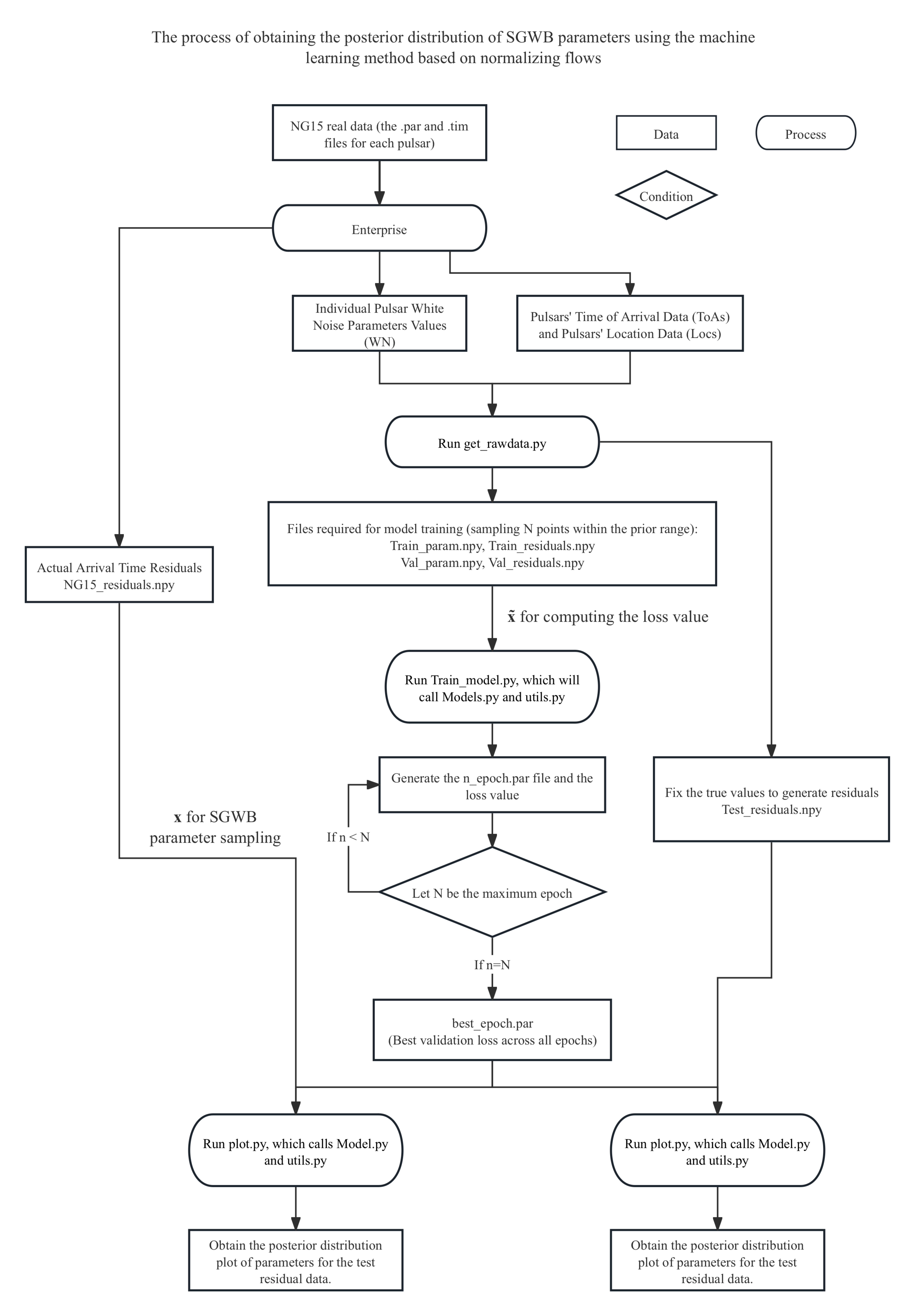}
    \caption{Workflow of the NF-based machine learning pipeline for SGWB analysis. The diagram outlines data extraction from the NG15 dataset, generation of simulated residuals, NF model training, and posterior inference.}
    \label{fig:processFIG}
\end{figure}

Figure~\ref{fig:processFIG} summarizes our normalizing flow (NF)-based machine learning pipeline for SGWB analysis. The workflow proceeds from the NG15 raw data to the final posterior distribution, enabling inference of 22 noise and SGWB parameters from pulsar timing residuals. The four key stages are:

\begin{enumerate}
    \item \textbf{Data Extraction:} Use \texttt{ENTERPRISE} to process the NG15 wideband dataset, obtaining pulsar sky positions, times of arrival (ToAs), white noise parameters, and true timing residuals.
    \item \textbf{Residual Generation:} Generate simulated datasets, SGWB+noise parameters and timing residuals.
    \item \textbf{NF Model Training:} Train the NF model on the simulated data using the architecture described in Ref.~\cite{PTAflow} and code from \cite{PTAflow} provided by Ref.~\cite{Shih:2023jme}.
    \item \textbf{Posterior Inference:} Feed the NG15 observational residuals into the trained NF model to obtain posterior distributions for the SGWB and noise parameters.
\end{enumerate}

\section{Descriptions for SGWB source models}
\label{sec:modeldescription}
\begin{enumerate}
    \item \textbf{Model 1:} Supermassive Black Hole Binaries (SMBHBs) with Environmental Effects (Bending Model). The SGWB spectrum from this model is given by \cite{QUINLAN199635,Phinney:2001di,NANOGrav:2015aud,2015PhRvD91h4055S}:
    \begin{equation}
    \Omega_{\mathrm{GW}}^{\mathrm{SMBHB}}(f) h^2 = \frac{2\pi^2}{3H_0^2} \, f^3 \, \frac{A^2}{12\pi^2} \, f_{\mathrm{yr}}^{\gamma-3} \, f^{-\gamma} \, \frac{1}{1 + \left(\frac{f_\mathrm{bend}}{f}\right)^{\kappa}},
    \end{equation}
    where \(A_{\mathrm{SMBHB}}\) is the amplitude of the SGWB produced by SMBHBs, and \(f_{\mathrm{bend}}\) is the frequency at which environmental effects (such as stellar hardening or gas interactions; here we consider stellar hardening, with \(\kappa = \frac{10}{3}\) \cite{NANOGrav:2015aud}) cause the spectrum to deviate from the canonical \(f^{2/3}\) power-law behavior, resulting in a spectral turnover.

    \item \textbf{Model 2:} Power-Law (PL) Model. The SGWB spectrum for this model is given by \cite{hobbs2009tempo2,2021MNRAS502478G}:
    \begin{equation}
    \Omega_{\mathrm{GW}}^{\mathrm{PL}}(f) h^2 = A_{\mathrm{PL}}^2 \, \frac{2 \pi^2}{3 H_0^2} \, f^{5-\gamma} \, f_\text{yr}^{\gamma-3} \, h^2,
    \label{eq:SMBH}
    \end{equation}
    where \(A_{\mathrm{PL}}\) denotes the amplitude of the power-law spectrum, \(\gamma\) is the spectral index that characterizes the frequency dependence, \(f_{\mathrm{yr}}=1\,\mathrm{yr}^{-1}\), and \(h\) is the dimensionless Hubble parameter.

    \item \textbf{Model 3:} Cosmic Strings (CS-META-L, Metastable Cosmic Strings). The SGWB spectrum from this model is given by \cite{Chang:2021afa,Auclair_2020,2021JCAP12006B,NANOGrav:2023hvm,Gouttenoire:2019kij}:
    \begin{equation}
    \Omega_{\mathrm{GW}}^{\mathrm{CS}}(f) h^2 = 
    \frac{8\pi (G\mu)^2}{3H_0^2} \sum_{k=1}^{k_{\mathrm{max}}} P_k \cdot I_k(f),
    \end{equation}
    where \(G\mu\) is the dimensionless string tension characterizing the energy scale of cosmic string formation, and $P_k = \frac{\Gamma}{\zeta(q)} \, \frac{1}{k^q}$
    represents the emission power of the \(k\)-th harmonic mode, with \(\Gamma\) and \(q\) being model parameters and \(\zeta(q)\) the Riemann zeta function. The frequency-dependent integral term is given by
    \begin{equation}
    I_k(f) = \frac{2k}{f} \int_{t_{\mathrm{ini}}}^{t_0} dt \, \left(\frac{a(t)}{a(t_0)}\right)^5 n_I\!\left(\frac{2k\,a(t)}{f\,a(t_0)}, t\right).
    \end{equation}
    For the metastable cosmic string (CS-meta) model, we have:
    $n_I^{\mathrm{meta}}(\ell, t) = \Theta(t_s - t_*) \, E(\ell, t) \, n_I(\ell, t)$, $t_s = \frac{1}{\Gamma_d^{1/2}}$, $t_* = \frac{\ell + \Gamma G\mu \, t}{\alpha_* + \Gamma G\mu}, \quad \alpha_* = \alpha(t_*)$,
    $\Gamma_d = \frac{\mu}{2\pi} e^{-\pi\kappa}$,
    $\sqrt{\kappa} = \frac{m_{\mathrm{GUT}}}{\mu^{1/2}} \sim \frac{\Lambda_{\mathrm{GUT}}}{\Lambda_{U(1)}}$, and $E(\ell, t) = e^{-\Gamma_d\left[\ell(t-t_*)+\frac{1}{2}\Gamma G\mu (t-t_*)^2\right]}$. The META-L metastable model assumes that cosmic strings are unstable to the formation of GUT monopoles and considers only the GW radiation from string loops. The decay parameter is characterized by \(\kappa\) (with \(\kappa \sim M_{\mathrm{GUT}}/\mu^{1/2}\)), where \(M_{\mathrm{GUT}}\) is the mass of the GUT gauge boson.

    \item \textbf{Model 4:} Domain Walls (DW). The SGWB spectrum for domain walls is given by \cite{Zhou:2020ojf,Hiramatsu:2013qaa,Kadota:2015dza}:
    \begin{equation}
        \Omega_{\mathrm{GW}}^{\mathrm{DW}}(f)h^2 = \Omega_{\mathrm{GW}}^{\mathrm{peak}} h^2 \, S^{\mathrm{dw}}(f),
    \end{equation}
    where the peak GW amplitude is:
    \begin{equation}
    \begin{aligned}
        \Omega_{\mathrm{GW}}^{\mathrm{peak}} h^2 \simeq & \,5.20 \times 10^{-20}\, \tilde{\epsilon}_{\mathrm{gw}} \, \mathcal{A}^4 \left(\frac{10.75}{g_*}\right)^{1/3} \\
        & \times \left(\frac{\sigma}{1\,\mathrm{TeV}^3}\right)^4 \left(\frac{1\,\mathrm{MeV}^4}{\Delta V}\right)^2,
    \end{aligned}
    \end{equation}
    and the shape function \(S^{\mathrm{dw}}(f)\) is defined as
    \begin{equation}
    \begin{aligned}
        S^{\mathrm{dw}}(f) &= \left(\frac{f}{f_{\mathrm{peak}}^{\mathrm{dw}}}\right)^3, \quad f < f_{\mathrm{peak}}^{\mathrm{dw}}, \\
        S^{\mathrm{dw}}(f) &= \left(\frac{f}{f_{\mathrm{peak}}^{\mathrm{dw}}}\right)^{-1}, \quad f \geq f_{\mathrm{peak}}^{\mathrm{dw}},
    \end{aligned}
    \end{equation}
    with the peak frequency estimated as \cite{Hiramatsu:2013qaa}:
    \begin{equation}
        f_{\mathrm{peak}}^{\mathrm{dw}} \simeq 3.99 \times 10^{-9}\,\mathrm{Hz}\,\mathcal{A}^{-1/2} \left(\frac{1\,\mathrm{TeV}^3}{\sigma}\right)^{1/2} \left(\frac{\Delta V}{1\,\mathrm{MeV}^4}\right)^{1/2}.
    \end{equation}
    Here, the prior parameters \(\sigma\) and \(\Delta V\) represent the domain wall tension and the bias potential that breaks the vacuum degeneracy, respectively. The bias potential causes the domain walls to decay and determines the position of the spectral peak. The area parameter is fixed to \(\mathcal{A} = 1.2\) \cite{Kadota:2015dza}, and the GW production efficiency is given by \(\tilde{\epsilon}_{\mathrm{gw}} = 0.7\) \cite{Zhou:2020ojf,Kadota:2015dza}.

    \item \textbf{Model 5:} First-Order Phase Transition (FOPT). The SGWB spectrum for FOPT is given by \cite{Caprini:2015zlo,Hirose:2023bvl,Zhou:2020ojf}:
    \begin{equation}
    \begin{aligned}
    \Omega_{\mathrm{GW}}^{\mathrm{FOPT}}(f) h^2 
    = 2.65 \times 10^{-6}\,&\left(H_* \tau_{\mathrm{sw}}\right) \left(\frac{\beta}{H_*}\right)^{-1} v_b \left(\frac{\kappa_v \alpha_{PT}}{1+\alpha_{PT}}\right)^2 \\
    &\times \left(\frac{g_*}{100}\right)^{-1/3} \left(\frac{f}{f_{\mathrm{peak}}^{\mathrm{FOPT}}}\right)^3 \left[\frac{7}{4+3\left(f/f_{\mathrm{peak}}^{\mathrm{FOPT}}\right)^2}\right]^{7/2},
    \end{aligned}
    \end{equation}
    with the peak frequency
    \begin{equation}
    f_{\mathrm{peak}}^{\mathrm{FOPT}} = 1.9 \times 10^{-5}\, \frac{\beta}{H_*}\, \frac{1}{v_b}\, \frac{T_*}{100}\, \left(\frac{g_*}{100}\right)^{1/6}\,\mathrm{Hz}.
    \end{equation}
    Here, \(\tau_{\mathrm{sw}} = \min\!\left[\frac{1}{H_*}, \frac{R_s}{U_f}\right]\) represents the duration of the sound wave phase, \(H_*\) is the Hubble parameter at temperature \(T_*\), and \(\alpha_{PT}\) (fixed at 1.0) quantifies the latent heat. Additionally, \(\beta/H_*\) characterizes the inverse duration of the phase transition, \(v_b\) is the bubble wall velocity (fixed at 0.975), and \(g_*\) is the effective number of relativistic degrees of freedom at the time of GW production.

    \item \textbf{Model 6:} Scalar Induced Gravitational Waves (SIGW-delta). The SGWB spectrum for this model is given by \cite{2019PhRvL122t1101C,2021JCAP10080A,2021PhLB82136606Y,2023PhRvD107d3520F,NANOGrav:2023hvm}:
    \begin{equation}
    \begin{aligned}
    \Omega_{\mathrm{GW}}^{\mathrm{SI}}(f) h^2 & = \frac{1}{12} \, \Omega_{\mathrm{rad}} h^2 \left(\frac{g_0}{g_*}\right)^{1/3} \\
    & \quad \times \int_0^{\infty} dv \int_{|1-v|}^{1+v} du \left(\frac{4v^2 - (1+v^2-u^2)^2}{4uv}\right)^2 \\
    & \quad \times P_{\mathcal{R}}(2\pi f u) \, P_{\mathcal{R}}(2\pi f v) \, I^2(u,v),
    \end{aligned}
    \end{equation}
    where
    \begin{equation}
    \begin{aligned}
    I^2(u,v) & = \frac{1}{2} \left(\frac{3}{4u^3 v^3 x}\right)^2 (u^2+v^2-3)^2 \\
    & \quad \times \Biggl\{ \left[-4uv + (u^2+v^2-3) \ln\!\left|\frac{3-(u+v)^2}{3-(u-v)^2}\right|\right]^2 \\
    & \quad + \left[\pi(u^2+v^2-3)\,\Theta(u+v-\sqrt{3})\right]^2 \Biggr\}.
    \end{aligned}
    \end{equation}
    This model describes the GW background generated at second order in perturbation theory by non-linear interactions of early-universe scalar perturbations. Here, \(\Omega_{\mathrm{rad}}\) is the present-day radiation energy density parameter, \(g_0\) and \(g_*\) are the effective relativistic degrees of freedom today and at the time of GW production, respectively, and \(I^2(u,v)\) is an integral kernel with complex dependencies. The SIGW-delta model is characterized by a delta-function form for the primordial curvature power spectrum:
    \begin{equation}
    P_{\mathcal{R}}(k) = \mathcal{A} \cdot \delta\!\left(\ln\!\left(\frac{k}{k_*}\right)\right),
    \end{equation}
    where \(\mathcal{A}\) is the amplitude of the perturbations, \(\delta\) is the Dirac delta function, and \(k_*\) is the characteristic wavenumber. This implies a sharply peaked scalar spectrum in logarithmic space, producing a significant GW signal at the corresponding characteristic frequency \(f_{\mathrm{peak}} = k_*/(2\pi a_0)\).

    \item \textbf{Model 7:}  Dual Scenario \((n_T, r)\)/Inflationary Gravitational Waves. This dual scenario describes a generalized inflationary and bouncing cosmology in the parameter space \((n_T, r)\). The SGWB spectrum is given by \cite{1974ZhETF..67..825G,1979JETPL..30..682S,1982PhLB..115..189R,1983PhLB..125..445F,1984NuPhB.244..541A, Caprini:2018mtu, NANOGrav:2023hvm,Li:2024oru}:
    \begin{equation}\label{eq:IGWspectrum}
    \Omega_{\mathrm{GW}}(f) h^2 = \frac{3}{128} \, \Omega_{\gamma 0} h^2 \, r \, P_R \left(\frac{f}{f_\ast}\right)^{n_T} \left[\left(\frac{f_{\mathrm{eq}}}{f}\right)^2 + \frac{16}{9}\right],
    \end{equation}
    where \(r\) and \(n_T\) are the tensor-to-scalar ratio and the spectral index of the primordial tensor spectrum, respectively. \(\mathcal{P}_R = 2 \times 10^{-9}\) is the amplitude of the curvature perturbation spectrum at the pivot scale \(k_\ast = 0.05\,\mathrm{Mpc}^{-1}\). \(f_\ast = 0.78\times 10^{-16}\,\mathrm{Hz}\) is the frequency today corresponding to \(k_\ast\), and \(f_\mathrm{eq} = 2.01\times 10^{-17}\,\mathrm{Hz}\) is the frequency today corresponding to matter–radiation equality. \(\Omega_{\gamma 0} = 2.474 \times 10^{-5}\,h^{-2}\) denotes the present-day radiation energy density fraction, and \(h = 0.677\) is the reduced Hubble constant.

    \item \textbf{Model 8:} Dual Scenario in the \((w, r)\) Plane. In this model, the SGWB spectrum is expressed in the parameter space \((w, r)\) and is given by \cite{Li:2024oru}:
    \begin{equation}
    \Omega_{\mathrm{GW}}(f) h^2 = \frac{3}{128} \, \Omega_{r0} h^2 \, r \, P_R \left(\frac{f}{f_\ast}\right)^{\frac{4}{3w+1}+2} \left[\left(\frac{f_{\mathrm{eq}}}{f}\right)^2 + \frac{16}{9}\right].
    \end{equation}
    where \(w\) is the equation of state (EoS) of inflation or bouncing cosmic background.

    \item \textbf{Model 9:} Dual Scenario with a Time-Independent (Stable) Scale-Invariant Solution in the \((m, r)\) Plane. The SGWB spectrum for this model is given by \cite{Li:2024oru}:
    \begin{equation}
    \Omega_{\mathrm{GW}}(f) h^2 = \frac{3}{128} \, \Omega_{r0} h^2 \, r \, P_R \left(\frac{f}{f_\ast}\right)^{-\frac{1}{2}m+1} \left[\left(\frac{f_{\mathrm{eq}}}{f}\right)^2 + \frac{16}{9}\right],
    \end{equation}
    where \(m\) is the modified damping parameter of  primordial curvature perturbation. 
    
    \item \textbf{Model 10:} Dual Scenario with a Time-Dependent (Dynamic) Scale-Invariant Solution in the \((m, r)\) Plane. The SGWB spectrum for this model is given by \cite{Li:2024oru}:
    \begin{equation}
    \Omega_{\mathrm{GW}}(f) h^2 = \frac{3}{128} \, \Omega_{r0} h^2 \, r \, P_R \left(\frac{f}{f_\ast}\right)^{\frac{1}{4}m+1} \left[\left(\frac{f_{\mathrm{eq}}}{f}\right)^2 + \frac{16}{9}\right].
    \end{equation}
\end{enumerate}

\section{Prior}
\label{sec:prior}
\begin{table}[h!]
\centering
\footnotesize
\begin{tabular}{l c c}
\toprule
\textbf{Parameter} & \textbf{Description} & \textbf{Prior} \\
\hline
\multicolumn{3}{c}{\textbf{Red Noise}} \\
$A_{\mathrm{RN}}$ & Red noise amplitude & log-uniform $[-19, -13]$ \\
$\gamma$ & Red noise spectral index & uniform $[1, 7]$ \\
\bottomrule
\end{tabular}
\caption{Prior ranges for red noise parameters. (Note: All logarithms are base 10.)}\label{tab:prior_rn}
\end{table}

\begin{table}[h!]
\hspace*{-1.2cm}
\centering
\footnotesize
\begin{tabular}{l c c}
\toprule
\textbf{Parameter} & \textbf{Description} & \textbf{Prior} \\
\hline
\multicolumn{3}{c}{\textbf{SMBHBs with Environment (Turnover Model)}} \\
$A_{\mathrm{SMBHB}}$ & SMBHBs amplitude & log-uniform $[-18, -12]$ \\
$f_{\mathrm{bend}}\mathrm{[Hz]}$ & Bending frequency & log-uniform $[-10, -7]$ \\
\hline
\multicolumn{3}{c}{\textbf{Powerlaw}} \\
$A_{\mathrm{PL}}$ & Powerlaw amplitude & log-uniform $[-18, -13]$ \\
$\gamma$ & Powerlaw spectral index & uniform $[1, 7]$ \\
\hline
\multicolumn{3}{c}{\textbf{Cosmic String(CS-metastable)}} \\
$G_{\mu}$ & String tension & log-uniform $[-14, -1.5]$ \\
$\sqrt{\kappa}$ & Decay parameter & uniform $[7, 9.5]$ \\
\hline
\multicolumn{3}{c}{\textbf{Domain Walls(DW)}} \\
$\sigma$ & Surface energy density & log-uniform $[0, 8]$ \\
$\Delta V$ & Bias potential & log-uniform $[0, 8]$ \\
\hline
\multicolumn{3}{c}{\textbf{First-order Phase Transitions(FOPT)}} \\
$\beta / H_{\star}$ & Inverse PT duration & uniform $[5, 70]$ \\
$T_\star\mathrm{[MeV]}$ & PT temperature & uniform $[0.01, 1.6]$ \\
\hline
\multicolumn{3}{c}{\textbf{Scalar-induced GWs(SIGW-delta)}} \\
$\mathcal{P}$ & Scalar amplitude & log-uniform $[-3, 1]$ \\
$f_{\mathrm{peak}}\mathrm{[Hz]}$ & Peak frequency & log-uniform $[-11, -5]$ \\
\hline
\multicolumn{3}{c}{\textbf{Dual scenario} ($n_T$, $r$)/IGW} \\
$n_T$ & Spectral index of the tensor spectrum & uniform $[-1, 6]$ \\
$r$ & Tensor-to-scalar ratio & log-uniform $[-16, 0]$ \\
\hline
\multicolumn{3}{c}{\textbf{Dual scenario} ($w$, $r$)} \\
$w$ & Equation of state parameter & uniform $[-10, 10]$ \\
$r$ & Tensor-to-scalar ratio & log-uniform $[-16, 0]$ \\
\hline
\multicolumn{3}{c}{\textbf{Stable Scale-invariant} ($m$, $r$)} \\
$m$ & Stable scale-invariant factor & uniform $[-32, 32]$ \\
$r$ & Tensor-to-scalar ratio & log-uniform $[-16, 0]$ \\
\hline
\multicolumn{3}{c}{\textbf{Dynamic Scale-invariant} ($m$, $r$)} \\
$m$ & Dynamic scale-invariant factor & uniform $[-32, 32]$ \\
$r$ & Tensor-to-scalar ratio & log-uniform $[-16, 0]$ \\
\bottomrule
\end{tabular}
\caption{Prior ranges for SGWB source parameters. (Note: All logarithms are base 10.)}\label{tab:prior_sgwb}
\end{table}

\section{Differet epoches}
\label{sec:epochs}
\begin{figure}[h!]
    \centering
    \includegraphics[width=1.\linewidth]{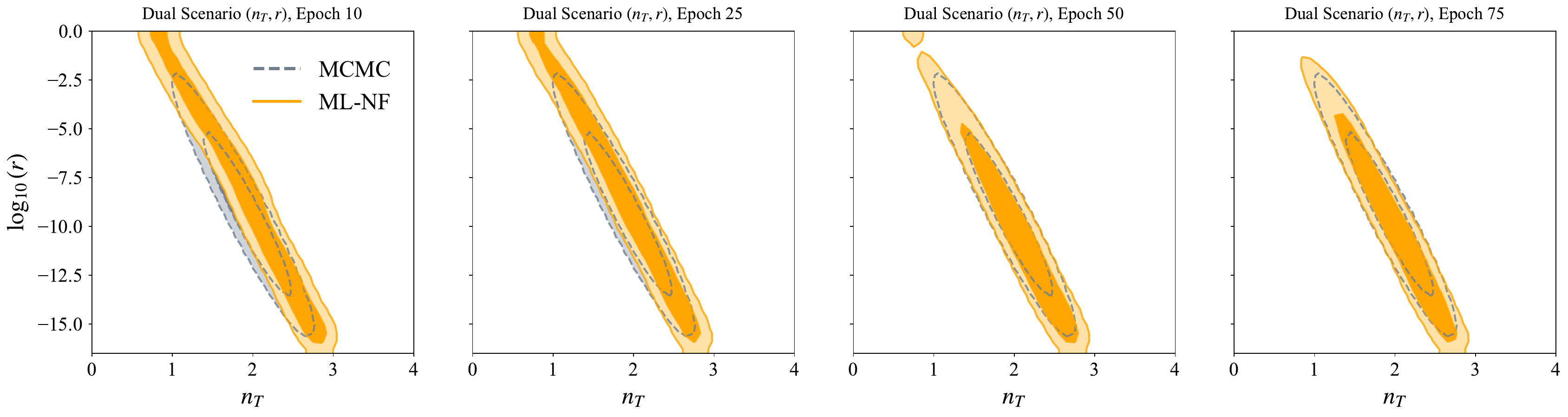}
    \caption{Comparison of training results at different epochs($n=10,25,50,75$).}
    \label{fig:diff_epochs}
\end{figure}

Through multiple training iterations, 
Fig.~\ref{fig:diff_epochs} compares the training results at different epochs for the Dual Scenario \((n_T, r)\) model using the NF-based ML method. For our purpose, training achieves sufficiently good performance by 50 epochs.

\section{Computational Workflow Comparison}  
\label{sec:time_comparison}

\begin{figure}[h!]
    \centering
    \includegraphics[width=0.9\linewidth]{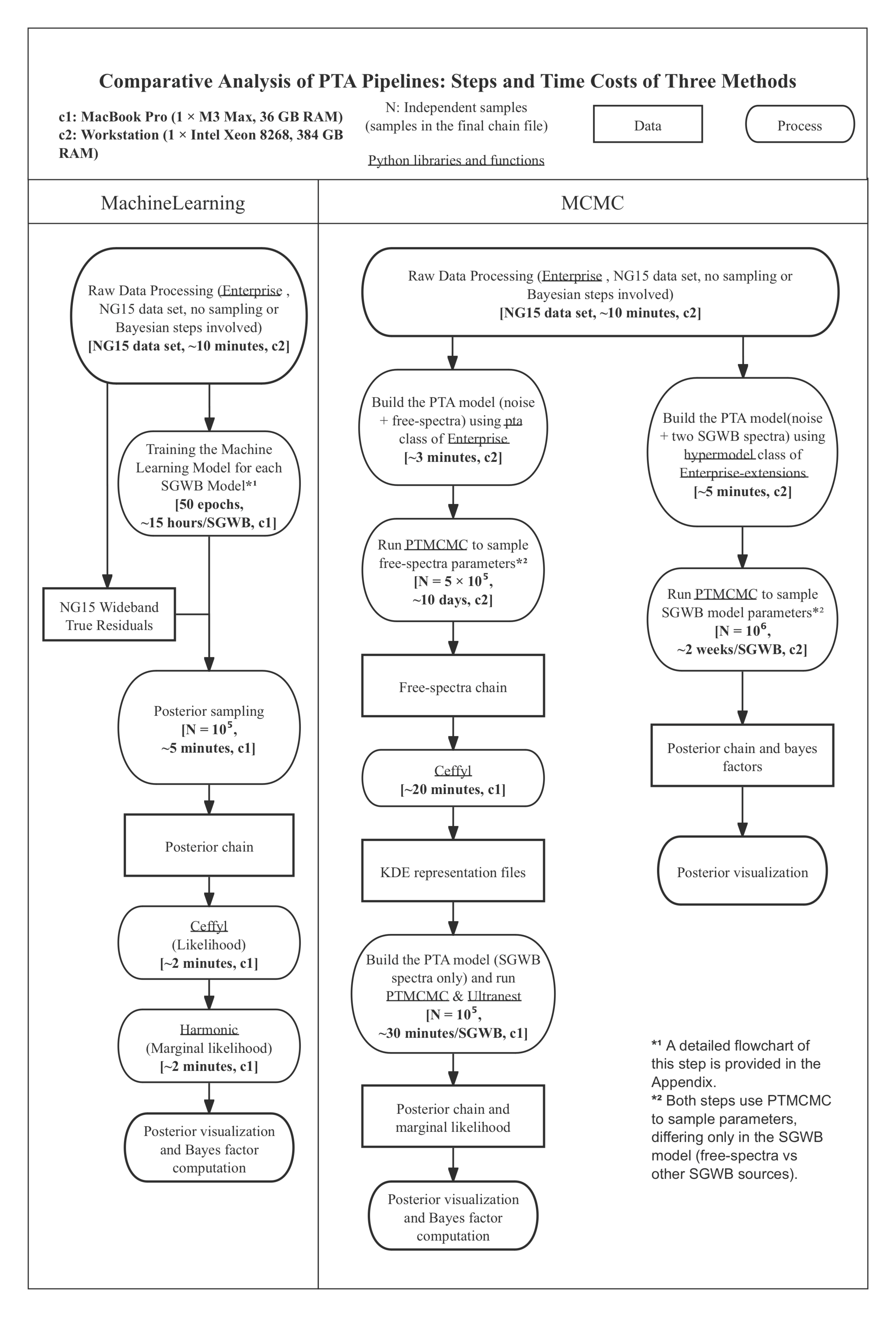}
    \caption{Comparative timing of three Bayesian analysis methods for PTA data.  }
    \label{fig:process-timeFIG}
\end{figure}

Fig.~\ref{fig:process-timeFIG} illustrates the comparative timing of three Bayesian analysis methods for PTA data. 

{\bf Left: Normalizing flow (NF)-based machine learning (ML-NF) workflow:}
   \begin{enumerate}
        \item Process NG15 raw data with \texttt{Enterprise}  
        \item Train ML model per SGWB scenario (training details in Appendix)  
        \item Perform posterior sampling with trained NF  
        \item Compute likelihoods via \texttt{ceffyl} and estimate marginal likelihoods  
        \item Visualize posteriors and calculate Bayes factors  
    \end{enumerate}
{\bf Right: MCMC approaches:}  

    \textbf{Method 1}:  
    \begin{enumerate}
        \item Build PTA model with \texttt{Enterprise}  
        \item Sample free spectrum (SGWB model) via PTMCMC  
        \texttt{Ultranest}-assisted SGWB parameter sampling  
        \item Compute posteriors and Bayes factors  
    \end{enumerate}
    
    \textbf{Method 2}:  
    \begin{enumerate}
        \item Construct PTA model with \texttt{Enterprise\_extensions}  
        \item Directly sample competing SGWB models via PTMCMC  
        \item Calculate Bayes factors from chains 
    \end{enumerate}
    Time estimates reflect full analysis cycles from raw data to visualization.

\section{Reweighted NF results}
\label{sec:reweighted}

To achieve more accurate posterior estimates for SGWB sources, the samples directly generated by the NF method can be reweighted using the likelihood $\mathcal{L}(\mathbf{x}_{\mathrm{obs}} \mid \theta^{(j)}_{(D=2)_l,\mathrm{SGWB}}, \mathcal{H}^{(j)}) $ \cite{Hourihane:2022ner, Dax:2022pxd, Shih:2023jme}, 
\begin{equation}\label{eq:reweighting}
    w_l^{(j)} = \frac{\mathcal{L}(\mathbf{x}_{\mathrm{obs}} \mid \theta^{(j)}_{(D=2)_l,\mathrm{SGWB}}, \mathcal{H}^{(j)}) \, \pi(\theta^{(j)}_l\mid \mathcal{H}^{(j)})}{p_{\phi_{\mathrm{best}}}(\theta^{(j)}_{(D=22)_l} \mid \mathbf{x}_{\mathrm{obs}}, \mathcal{H}^{(j)})},
\end{equation}
where $\{w_l^{(j)}\}$ is reweighting parameter dataset and $\pi(\theta^{(j)}_l\mid \mathcal{H}^{(j)})$ is prior listed in Table~\ref{tab:prior_rn} and Table~\ref{tab:prior_sgwb}. Using \texttt{corner} package \cite{Corner} to upload posterior samples of SGWB source model, $p_{\phi_{\mathrm{best}}}(\theta^{(j)}_{(D=2)_l, SGWB} \mid \mathbf{x}_{\mathrm{obs}}, \mathcal{H}^{(j)})$, together with their weights $\{w_l^{(j)}\}$, we obtained the reweighted posterior distributions, $p_{\phi_{\mathrm{best}}}^{\mathrm{RW}}(\theta^{(j)}_{(D=2)_l, SGWB} \mid \mathbf{x}_{\mathrm{obs}}, \mathcal{H}^{(j)})$, as illustrated in Fig.~\ref{fig:A_fbend_RW.pdf}-Fig.~\ref{fig:r_m_RW.pdf}. 

\section{Hellinger Distance Comparison}
\label{sec:hellinger}
Let \(f(x)\) and \(g(x)\) be two probability density functions defined over an \(N\)-dimensional parameter space. Their squared Hellinger distance \(H^2\) is defined as \cite{Hellinger+1909+210+271,Lamb:2023jls}:
\begin{gather}
    H^2(f,g) = \int \left(\sqrt{f(x)} - \sqrt{g(x)}\right)^2 dx 
    = 1 - \int \sqrt{f(x)g(x)} \, dx,
    \label{eq:hd_def}
\end{gather}
which quantifies the similarity between the posterior samples of two different distributions. The Hellinger distance is bounded between 0 and 1, with smaller values indicating closer agreement between the distributions. In practice, \(H < 0.3\) implies that the two distributions are well aligned.

In this study, we let \(f(x)\) denote the (reweighted) NF-based posterior,
\(
p_{\phi_{\mathrm{best}}}^{(\mathrm{RW})}\!\left(\theta^{(j)}_{(D=2)_l,\mathrm{SGWB}} \mid \mathbf{x}_{\text{obs}}, \mathcal{H}^{(j)}\right),
\)
and \(g(x)\) denote the MCMC posterior,
\(
p_{\mathrm{MCMC}}\!\left(\theta^{(j)}_{(D=2)_l,\mathrm{SGWB}} \mid \mathbf{x}_{\text{obs}}, \mathcal{H}^{(j)}\right).
\)
These functions compare the (reweighted) NF-based posterior and the MCMC posterior, respectively, as presented in Table~\ref{Hellinger_distances}.

\section{Physical Interpretation of SGWB Source Model Comparison}
\label{sec:phyint}

Table~\ref{tab:bayes_factors_combined_new} summarizes the Bayes factor comparisons between SGWB source models. Both MCMC and NF evidence estimates indicate that the dual \((w,r)\) scenario is most strongly favored, with Bayes factors \(\gtrsim6\) against nearly every alternative and \(\gtrsim300\) relative to the domain wall model. The dual “stable” and “dynamic” scenarios follow closely, outperforming standard astrophysical models—such as SMBHBs, power‑law, and cosmic strings—by factors of a few and decisively beating domain walls (\(\mathrm{BF}\sim10^2\)). Scalar‑induced GWs and the pure power‑law model occupy a mid‑tier, with moderate support (\(\mathrm{BF}\sim1\text{–}3\)) over SMBHBs and cosmic strings but still \(\mathcal{O}(10^2)\) above domain walls. SMBHBs and the inflationary IGW model exhibit only weak to positive evidence relative to each other (\(\mathrm{BF}\sim1\text{–}2\)) and are modestly preferred over cosmic strings and first‑order phase transitions. First‑order phase transitions barely outscore domain walls (\(\mathrm{BF}\sim20\)), while domain walls remain the least favored hypothesis (\(\mathrm{BF}\ll1\) compared to any other model).

\end{appendix}
\bibliography{biblio}
\bibliographystyle{apsrev}

\end{document}